\documentclass[runningheads]{llncs}
\usepackage[T1]{fontenc}
\usepackage{graphicx}
\usepackage{changepage} 
\usepackage{tikz}
\usepackage{amsfonts}
\usepackage{mathtools}
\usepackage[symbol*]{footmisc}

\usepackage[colorlinks=true,citecolor=red]{hyperref}

\usepackage[space]{cite}

\newcommand{\da}{{\sc Defensive Alliance}}

\begin{document}
\title{On the Tractability of Defensive Alliance Problem\thanks{Preliminary version of this paper has appeared in \textit{9th Annual International Conference on
Algorithms and Discrete Applied Mathematics, CALDAM 2023}}}

\author{Sangam Balchandar Reddy \and
Anjeneya Swami Kare}
\authorrunning{Balchandar Reddy and A. S. Kare }
\titlerunning{Tractability of defensive alliance in graphs}
%
\institute{School of Computer and Information Sciences, University of Hyderabad,\\ Hyderabad, India.\\
\email{\{21mcpc14,askcs\}@uohyd.ac.in}}
\maketitle              
\begin{abstract}
Given a graph $G = (V, E)$, a non-empty set $S \subseteq V$ is a \da{}, if for every vertex $v \in S$,  the majority of its closed neighbours are in $S$, that is, $|N_G[v] \cap S| \geq |N_G[v] \setminus S|$. The decision version of the problem is known to be NP-Complete even when restricted to split and bipartite graphs. The problem is \textit{fixed-parameter tractable} for the parameters solution size, vertex cover number and neighbourhood diversity. For the parameters treewidth and feedback vertex set number, the problem is W[1]-hard. \\
\hspace*{2em} In this paper, we study the \da{} problem for graphs with bounded degree. We show that the problem is \textit{polynomial-time solvable} on graphs with maximum degree at most 5 and NP-Complete on graphs with maximum degree 6. This rules out the fixed-parameter tractability of the problem for the parameter maximum degree of the graph. We also consider the problem from the standpoint of parameterized complexity. We provide an FPT algorithm using the Integer Linear Programming approach for the parameter distance to clique. We also answer an open question posed in \cite{AG2} by providing an FPT algorithm for the parameter twin cover.

\end{abstract}
\section{Introduction}
We, humans, form alliances for the sake of mutual benefit. This can often be seen in politics, businesses, trades, etc. The main agenda behind the alliance is to achieve a common goal between the parties. Based on this idea, the alliances are classified as follows. An alliance is formed between the parties of a group to defend against an attack from other parties (defensive alliance) or to be able to attack other parties (offensive alliance). The concept of alliances in graphs was first introduced by Kristiansen, Hedetniemi and Hedetniemi~\cite{KRP}. The initial algorithmic results of the problem were given by Jamieson~\cite{LHJ2}. Alliances in graphs have been well studied~\cite{CM} and various generalizations are also studied~\cite{HF2,KHS}. \vspace{2mm} \\
\indent A defensive alliance is a non-empty set $S \subseteq V$, such that for every vertex $v \in S: |N_G[v] \cap S| \geq |N_G[v] \setminus S|$. We say that a vertex $v$ is protected if it has at least as many closed neighbours inside $S$ as it has outside $S$. The boundary of a set $S \subseteq V$ is denoted by $\partial S$ and represents the vertices in the neighbourhood of $S$, excluding $S$, i.e., $\partial S$ = $N[S]\setminus S$. An offensive alliance is a non-empty set $S \subseteq V$, such that for every vertex $v \in \partial S: |N_G[v] \cap S| \geq |N_G[v] \setminus S|$.  A powerful alliance is both defensive and offensive simultaneously. An alliance is global if it is also a dominating set. In this paper, we confine our study to the D\scriptsize{EFENSIVE ALLIANCE }\normalsize problem. We define the decision version of the problem as follows:  \vspace{3mm} \\
\noindent\fbox{%
    \parbox{\textwidth}{%
        D\scriptsize{EFENSIVE ALLIANCE:} \normalsize \\
        $\hspace*{0cm}$ \normalsize \textbf{Input:} A simple, undirected graph $G = (V, E)$, and a positive integer $k$. \\
        $\hspace*{0cm}$ \textbf{Question:} Is there a defensive alliance $S \subseteq V$ such that $|S| \leq k$?
    }%
} \vspace{2mm} \\
The optimization version of the problem asks to compute the defensive alliance with minimum cardinality. \vspace{3mm} \\
\textit{Known results.} There are polynomial time algorithms for finding minimum alliances\footnote{By alliances, we mean defensive, offensive, and powerful alliance problems} in trees~\cite{CHNG,LHJ2}. Kiyomi and Otachi~\cite{MK} have provided an XP algorithm on graphs with bounded treewidth. There is a polynomial time algorithm for finding a minimum defensive alliance in series-parallel graphs~\cite{RE}. Jamieson et al.~\cite{LHJ1} showed that the defensive alliance is NP-Complete even when restricted to split and bipartite graphs. Gaikwad and Maity~\cite{AG2} proved that the defensive alliance is NP-Complete on circle graphs. Fernau and Raible~\cite{HF1} proved that the alliance problems, including their global versions, are \textit{fixed-parameter tractable} when parameterized by the solution size. Enciso~\cite{RE} proved that defensive and global defensive alliances are \textit{fixed-parameter tractable} parameterized by domino treewidth. Alliances are \textit{fixed-parameter tractable} when parameterized by vertex cover number of the input graph~\cite{MK}. Recently, both the defensive and offensive alliances were also shown to be \textit{fixed-parameter tractable} parameterized by the neighborhood diversity of the input graph~\cite{AG1}. Defensive alliance is W[1]-hard parameterized by a wide range of parameters such as the feedback vertex set, treewidth, cliquewidth, treedepth and pathwidth~\cite{AG2}. \vspace{3mm} \\
\textit{Our results.} We investigate the complexity of the D\scriptsize{EFENSIVE ALLIANCE }\normalsize problem on graphs with bounded degree. We also study the fixed-parameter tractability of the problem for the parameters distance to clique and twin cover. Our main findings are as follows:
\begin{enumerate}
    \item We show that the D\scriptsize{EFENSIVE ALLIANCE }\normalsize problem is \textit{polynomial-time solvable} on graphs with maximum degree at most 5.
    \item We prove that the D\scriptsize{EFENSIVE ALLIANCE }\normalsize problem is NP-Complete on graphs with maximum degree 6. We give a reduction from the well-known NP-Complete problem, D\scriptsize{OMINATING SET }\normalsize on cubic graphs.
    \item We also show that the D\scriptsize{EFENSIVE ALLIANCE }\normalsize problem is FPT parameterized by distance to clique.
    \item We provide an FPT algorithm for the D\scriptsize{EFENSIVE ALLIANCE }\normalsize problem parameterized by twin cover, which answers an open question posed in~\cite{AG2}.
\end{enumerate}
\section{Preliminaries}
\textit{Notation and terminology.}
We consider only simple, finite, connected and undirected graphs. Let $G = (V, E)$ be a graph with $V$ as the vertex set and $E$ as the edge set such that $n = |V|$ and $m = |E|$. $\Delta(G)$ represents the maximum degree of $G$. We denote the open neighbourhood of a vertex $v$ by $N_G(v)$ and the closed neighbourhood by $N_G[v]$. The set of vertices that belong to $N_G(v)$ and $N_G[v]$, respectively, are referred to as the neighbours and closed neighbours of a vertex $v$. The open neighbourhood of a set $S$ is denoted by $N_G(S)$ and the closed neighbourhood by $N_G[S]$. $N_G(S)$ = $\bigcup\limits_{v \in S}^{} N_G(v)$ and $N_G[S]$ = $\bigcup\limits_{v \in S}^{} N_G[v]$. The degree of a vertex $v$ is represented by $d(v)$ and $d(v) = |N(v)|$. $dist(a, b)$ represents the shortest-path distance between the vertices $a$ and $b$. $SP(a, b)$ represents the set of vertices in the shortest path between $a$, $b$ including $a$, $b$. The girth of a graph $G$, is the length of the shortest cycle in $G$ and is denoted by $g$. A graph is $r$-regular if each vertex in the graph has a degree exactly $r$. A cubic graph is a 3-regular graph. \vspace{3mm} \\
\textit{Parameterized Complexity.} A problem is considered to be \textit{fixed-parameter tractable} w.r.t. a parameter $k$, if there exists an algorithm with running time $\mathcal{O}(f(k) \cdot n^{\mathcal{O}(1)})$, where $f$ is a computable function. We use $\mathcal{O}^*(f(k))$ to denote the time complexity of the form $\mathcal{O}(f(k) \cdot n^{\mathcal{O}(1)})$. Similarly, the term W[1]-hard is used to express the hardness of a problem w.r.t. a parameter. In parameterized complexity, the hierarchy of complexity classes is defined as follows: FPT $\subseteq$ W[1] $\subseteq$ W[2] $\subseteq$ ... $\subseteq$ XP. In general, FPT $\neq$ W[1] under the Exponential Time Hypothesis~\cite{ETH}. Class W[1] is the analog of NP in parameterized complexity. For more information on \textit{graph theory} and \textit{parameterized complexity}, we refer the reader to~\cite{WEST} and~\cite{MC}, respectively.
\section{D\footnotesize{EFENSIVE ALLIANCE }\normalsize on graphs with maximum degree at most 5}
In this section, we show that the D\scriptsize{EFENSIVE ALLIANCE }\normalsize problem is \textit{polynomial-time solvable} on graphs with maximum degree at most 5. For the rest of the section, we assume that $\Delta(G) \leq 5$.\vspace{3mm} \\
\textbf{Theorem 1.} \textit{The} D\scriptsize{EFENSIVE ALLIANCE }\normalsize  \textit{problem on graphs with $\Delta(G) \leq 5$  is \textit{polynomial-time solvable}}. \vspace{3mm} \\
\textbf{Lemma 1.} A set containing only the vertices of a cycle in $G$, forms a defensive alliance. \vspace{3mm} \\
\textit{Proof.} Let $S$ be a set containing only the vertices of a cycle. All the vertices in $S$ will have three closed neighbours in $S$ and at most three neighbours outside $S$. Therefore, all the vertices in $S$ are protected, making it a defensive alliance. \qed \vspace{3mm}
\setlength{\textfloatsep}{1\baselineskip plus 0\baselineskip minus 0\baselineskip}
\begin{figure} 
    \centering
    \begin{tikzpicture} [scale = 0.5]
    \draw[black] (1, 0) circle(0.15);
    \draw(2.5, 1) circle(0.15);
    \filldraw[red] (4, 0) circle(0.15);
    \draw(2.5, -1) circle(0.15);
    \filldraw[blue](6, 0) circle(0.15);
    \filldraw[blue](8, 1) circle(0.15);
    \filldraw[blue](8, -1) circle(0.15);
    \draw(10, 1) circle(0.15);
    \draw(10, -1) circle(0.15);
    \draw[thin, black] (1.1, 0.1) -- (2.4, 0.9);
    \draw[thin, black] (1.1, -0.1) -- (2.4, -0.9);
    \draw[thin, black] (3.9, 0.1) -- (2.6, 0.9);
    \draw[thin, black] (3.9, -0.1) -- (2.6, -0.9);
    \draw[thin, black] (4.15, 0) -- (5.85, 0);
    \draw[thin, black] (6.1, 0.1) -- (7.9, 0.9);
    \draw[thin, black] (6.1, -0.1) -- (7.9, -0.9);
    \draw[thin, black] (8, 0.85) -- (8, -0.85);
    \draw[thin, black] (6, 0) -- (9.88, 0.92);
    \draw[thin, black] (6, 0) -- (9.88, -0.92);
    \draw[thin, black] (8, 1) -- (9.85, 1);
    \draw[thin, black] (8, 1) -- (9.9, -0.9);
    \draw[thin, black] (8, -1) -- (9.9, 0.9);
    \draw[thin, black] (8, -1) -- (9.85, -1);
    \draw[thin, black] (10, -0.85) -- (10, 0.85);
    \filldraw (0.8,0.05) circle (0cm) node[anchor=south]{$v_1$};
    \filldraw (2.35,1.1) circle (0cm) node[anchor=south]{$v_2$};
    \filldraw (4,0.15) circle (0cm) node[anchor=south]{$v_4$};
    \filldraw (2.35,-1.8) circle (0cm) node[anchor=south]{$v_3$};
    \filldraw (6,0.15) circle (0cm) node[anchor=south]{$v_5$};
    \filldraw (8.25,1.1) circle (0cm) node[anchor=south]{$v_6$};
    \filldraw (8.25,-1.8) circle (0cm) node[anchor=south]{$v_7$};
    \filldraw (10.25,1.1) circle (0cm) node[anchor=south]{$v_8$};
    \filldraw (10.25,-1.8) circle (0cm) node[anchor=south]{$v_9$};

    \draw[black] (14, 0) circle(0.15);
    \draw(15.5, 1) circle(0.15);
    \draw[black] (17, 0) circle(0.15);
    \draw[black](15.5, -1) circle(0.15);
    \filldraw[red](19, 0) circle(0.15);
    \filldraw[blue](21, 1) circle(0.15);
    \filldraw[blue](21, -1) circle(0.15);    
    \draw(23, 1) circle(0.15);
    \draw(23, -1) circle(0.15);
    \draw[thin, black] (14.1, 0.1) -- (15.4, 0.9);
    \draw[thin, black] (14.1, -0.1) -- (15.4, -0.9);
    \draw[thin, black] (16.9, 0.1) -- (15.6, 0.9);
    \draw[thin, black] (16.9, -0.1) -- (15.6, -0.9);
    \draw[thin, black] (17.15, 0) -- (18.85, 0);
    \draw[thin, black] (19.1, 0.1) -- (20.9, 0.9);
    \draw[thin, black] (19.1, -0.1) -- (20.9, -0.9);
    \draw[thin, black] (21, 0.85) -- (21, -0.85);
    \draw[thin, black] (19.14, 0.05) -- (22.88, 0.92);
    \draw[thin, black] (19.14, -0.05) -- (22.88, -0.92);
    \draw[thin, black] (21, 1) -- (22.85, 1);
    \draw[thin, black] (21, 1) -- (22.9, -0.9);
    \draw[thin, black] (21, -1) -- (22.9, 0.9);
    \draw[thin, black] (21, -1) -- (22.85, -1);
    \draw[thin, black] (23, -0.85) -- (23, 0.85);
    \filldraw (13.8,0.05) circle (0cm) node[anchor=south]{$v_1$};
    \filldraw (15.35,1.1) circle (0cm) node[anchor=south]{$v_2$};
    \filldraw (17,0.15) circle (0cm) node[anchor=south]{$v_4$};
    \filldraw (15.35,-1.8) circle (0cm) node[anchor=south]{$v_3$};
    \filldraw (19,0.15) circle (0cm) node[anchor=south]{$v_5$};
    \filldraw (21.25,1.1) circle (0cm) node[anchor=south]{$v_6$};
    \filldraw (21.25,-1.8) circle (0cm) node[anchor=south]{$v_7$};
    \filldraw (23.25,1.1) circle (0cm) node[anchor=south]{$v_8$};
    \filldraw (23.25,-1.8) circle (0cm) node[anchor=south]{$v_9$};
    \end{tikzpicture} \begin{adjustwidth}{0pt}{0pt}
    \caption{Illustration of \textit{Observation 1}, with an instance of the subproblem $P_4$ on the left and an instance of the subproblem $P_5$ on the right.
    }
    \label{fig:Fig1}
    \end{adjustwidth}
\end{figure}
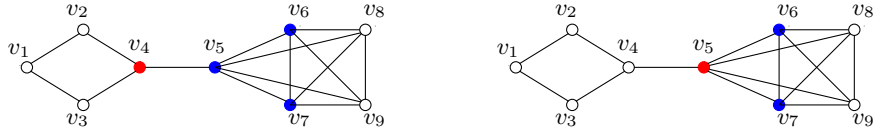
Consider a graph $G = (V, E)$, where $|V| = n$ and $V = \{v_1, v_2, ..., v_n\}$. We solve $n$ independent subproblems: $P_1, P_2, ..., P_n$, where $P_i$ denotes the problem of computing the smallest defensive alliance of $G$ containing the vertex $v_i$. \vspace{3mm} \\
\textbf{Observation 1:} Consider the graph shown in \autoref{fig:Fig1}. The union of the vertices of the path $v_4$, $v_5$ and the cycle $v_5, v_6$ and $v_7$ forms a defensive alliance for the subproblem $P_4$ which includes the vertex $v_4$. However, the cycle $v_5, v_6$ and $v_7$ forms a defensive alliance of smaller cardinality in subproblem $P_5$. \vspace{3mm} \\
\indent From observation 1, it is clear that in order to compute the optimal solution to the original problem, we can ignore a case where we consider the vertices of a cycle and the path joining the cycle to obtain the defensive alliance for subproblems. For the rest of this section, we compute the solutions to the subproblems by avoiding such a scenario. \vspace{3mm} \\
\textbf{Observation 2: }If $d(v_i) \leq 1$, then $v_i$ itself forms a defensive alliance. So, we consider only the subproblems in which $d(v_i) \geq 2$. \vspace{3mm} \\
\noindent\textbf{Lemma 2.} Let $d(v_i) = 3$ [or $2$], the size of the optimal solution for $P_i$, will be minimum among the following:
\begin{enumerate}
    \item $\min_{x \in W} dist(v_i, x)$+1, where $W = \{u \in V(G) \setminus v_i | d(u) \leq 3\}$.
    \item length of the shortest cycle containing $v_i$. 
\end{enumerate}
\textit{Proof.} Let $u_1, u_2$ and $u_3$ [or $u_1$ and $u_2$] are the neighbours of $v_i$. \vspace{1mm} \\
Case 1: Let us assume that exactly one neighbour of $v_i$ (say $u_1$) is picked into the solution. As neither $u_2$ nor $u_3$ [or $u_2$] can be a part of the solution, it is clear that the defensive alliance cannot be a set of vertices from a cycle containing $v_i$. We cannot add any more vertices that would form a cycle outside $v_i$, as explained earlier in Observation 1. Then the only way we could form a defensive alliance containing $v_i$ and $u_1$ is to find the closest vertex (say $x$) from $v_i$, with $d(x) \leq 3$ and add all the vertices in the shortest path joining $u_1$ and $x$ to the set. Except $x$ and $v_i$ all the other vertices of the path has at least three closed neighbours in the set and both $x_i$ and $v_i$ with the degree of at most three has at least two closed neighbours in the set. Hence, the vertices of the path joining $x$ and $v_i$ forms a defensive alliance. If no such $x$ exists, then $P_i$ cannot lead to an optimal solution.\vspace{1mm} \\
Case 2: Let us assume that exactly two neighbours of $v_i$ (say $u_1$ and $u_2$) are picked into the solution. The only way to form a defensive alliance set containing the vertices $v_i, u_1$ and $u_2$ is to find the shortest cycle containing $v_i, u_1$ and $u_2$ and add all the vertices in the cycle to the set. If no such cycle exists, then $P_i$ cannot lead to an optimal solution.\vspace{1mm} \\
Case 3: Let us assume that all three neighbours of $v_i$ are picked into the solution. If the defensive alliance forms a cycle containing the three neighbours of $v_i$ then we can infer that there is a smaller cycle that contains just the two neighbours of $v_i$. Hence, this case would lead to a larger set than either case 1 or case 2. \vspace{2mm} \\
\indent We will have three [or two] combinations in case 1 and three [or one] combinations in case 2. The minimum among them will return the  optimal solution for the subproblem $P_i$. \qed \vspace{2mm}
\noindent\textbf{Lemma 3.} Let $d(v_i) = 5$ [or $4$], the size of the optimal solution for $P_i$ will be the minimum among the following:
\begin{enumerate}
    \item $\min_{x \neq y \in W} (dist(v_i, x) + dist(v_i, y)+1$), $W = \{u \in V(G) \setminus v_i | d(u) \leq 3\},  \{SP(v_i,x) \cap SP(v_i, y)\} \setminus v_i = \emptyset$.
    \item length of the shortest cycle containing $v_i$.
\end{enumerate}
\textit{Proof.} Let $u_1, u_2, u_3, u_4$ and $u_5$ [or $u_1, u_2, u_3$ and $u_4$] are the neighbours of $v_i$.\\
Case 1: Let us assume that exactly one neighbour of $v_i$ (say $u_1$) is picked into the solution. Then, it is easy to see that $v_i$ is not protected. Therefore, no defensive alliance is possible by picking only one neighbour of $v_i$ as part of the solution.\\
Case 2: Let us assume that exactly two neighbours of $v_i$ (say $u_1$ and $u_2$) are picked into the solution. The defensive alliance containing the vertices $v_i, u_1$ and $u_2$ can be formed either with the vertices of the shortest cycle containing $v_i, u_1$ and $u_2$ or with the vertices along the two vertex disjoint shortest paths from $u_1$ and $u_2$ to vertices with degrees at most three. The minimum among the two sets forms a defensive alliance.\\
Case 3: Let us assume that more than two neighbours of $v_i$ are picked into the solution, then as explained in case 3 of Lemma 2, this forms a defensive alliance with a larger cardinality than case 2. \vspace{2mm} \\
\indent We will have a total of twenty [or twelve] combinations in case 2. The minimum among them will give the optimal solution for the subproblem $P_i$. \qed \vspace{3mm}
\noindent Given a vertex $v$,
\begin{enumerate}
\item Computing the closest vertex to $v$ with degree at most three can be done in $O(m+n)$ time.
\item Using BFS, we can compute the shortest cycle containing $v$ in $O(m+n)$ time.
\item Finding two vertex disjoint shortest paths to vertices with degree at most three from $v$, is also solvable in $O(m+n)$ time.
\end{enumerate}
Therefore, we can compute the optimal solution (if it exists) of the subproblem $P_i$ in linear time. Hence, the D\scriptsize{EFENSIVE ALLIANCE }\normalsize problem on graphs with $\Delta(G) \leq 5$ is polynomial-time solvable. This concludes the proof of Theorem 1.
\section{D\footnotesize{EFENSIVE ALLIANCE }\normalsize on graphs with maximum degree 6}
In this section, we prove that the D\scriptsize{EFENSIVE ALLIANCE }\normalsize problem is NP-Complete on graphs with maximum degree 6. \vspace{3mm} \\
\noindent \textbf{Theorem 2.} D\scriptsize{EFENSIVE ALLIANCE }\normalsize \textit{on graphs with $\Delta(G) = 6$ is NP-Complete.}\vspace{2mm}\\
It is easy to see that the problem is in NP. To prove the NP-Hardness, we reduce from the following problem: \vspace{3mm} \\
\noindent\fbox{%
    \parbox{\textwidth}{%
        D\scriptsize{OMINATING SET ON CUBIC GRAPHS:} \normalsize \vspace{1mm} \\
        Given a cubic graph $G = (V, E)$, a set $S \subseteq V$ is a dominating set if every vertex $v \in V\setminus S$ has a neighbour in $S$.\vspace{2mm} \\
        \noindent
        $\hspace*{0cm}$ \normalsize \textbf{Input:} A simple, undirected cubic graph $G = (V, E)$, and a positive integer $k$. \\
        $\hspace*{0cm}$ \textbf{Question:} Is there a dominating set $S \subseteq V$ such that $|S| \leq k$?
    }%
} \vspace{3mm} \\
In 1980, Kikuno et al. \cite{Kikuno} proved that the D\scriptsize{OMINATING SET} \normalsize problem on cubic graphs is NP-Complete. \vspace{3mm} \\
Given an instance $I = (G, k)$ of the D\scriptsize{OMINATING SET }\normalsize problem with $G$ being a cubic graph, we construct an instance of $I' = (G', k')$ of the D\scriptsize{EFENSIVE ALLIANCE }\normalsize problem. We need a special type of vertices in $G'$ that cannot be a part of any defensive alliance of size at most $k'$, we call them the forbidden vertices. The forbidden vertices are indicated using square-shaped nodes. We make use of the following gadget to generate the forbidden vertices. \vspace{3mm} \\
\noindent \textbf{Gadget to generate the forbidden vertices:} \vspace{3mm}\\
 Consider a 6-regular graph. Note that all the vertices have a closed neighbourhood of seven. To protect a vertex of the defensive alliance from a 6-regular graph, it should have at least four closed neighbours in the set. Therefore, the optimal defensive alliance can be obtained by finding the \textit{minimum induced subgraph of minimum degree $\geq $ 3}(from here on referred to as MSMD(3)). As we do not want any vertex from the gadget to be a part of the defensive alliance, we construct a 6-regular graph with no MSMD(3) of size at most $k'$. We make use of Ramanujan graphs to construct the gadget. \vspace{2mm} \\
 \textbf{Definition 1.} A \textit{Ramanujan graph} is a $r$-regular graph whose non-trivial eigenvalues lie in the interval $[-2\sqrt{r-1}, 2\sqrt{r-1}]$. For more information on Ramanujan graphs, we refer the reader to \cite{MORGENSTERN, Ramanujan}.\vspace{2mm} \\
From \cite{Ramanujan}, we have that $r$-regular Ramanujan graphs have girth, $g \geq \frac{4}{3} \log_{r-1} |V|$. \vspace{2mm} \\
Given the minimum degree $r$ and girth $g$, we have the following lower bound on the graph size $|V(r, g)|$. \vspace{2mm} \\
\indent $|V(r, g)| \geq \frac{r(r-1)^{\frac{g-1}{2}}-2}{r-2}$ for odd $g$\\
\indent $|V(r, g)| \geq \frac{2(r-1)^{\frac{g}{2}}-2}{r-2}$ for even $g$ \vspace{4mm} \\
\textbf{Lemma 4.} The size of the 6-regular Ramanujan graph that has an MSMD(3) of size $k'+1$ is polynomial in $k'$. \vspace{2mm} \\
\textit{Proof.} Consider a 6-regular Ramanujan graph $R$. $R$ has girth at least $\frac{4}{3} \log_5 n$. Let $M$ be an MSMD(3) of $R$. As $M$ is an induced subgraph of $R$, it also has a girth of at least $\frac{4}{3} \log_5 n$. From the lower bound on the size of the graph $|V(r, g)|$, we have that the size of $M$ with girth at least $\frac{4}{3} \log_5 n$ is $\Omega(2^{\frac{\frac{4}{3} \log_5 n}{2}})$ which is $\Omega(n^{\frac{2}{3}\log_5 2})$. The bound can be represented as $\Omega(n^c)$ where $c = 0.2871$. 

$k'+1 \ge c_1 \cdot 2^{\frac{\frac{4}{3}\log_5 n}{2}}$ $\implies$
$k'+1 \ge c_1 \cdot n^{\frac{2}{3}\log_5 2}$ $\implies$
$k'+1 \ge c_1 \cdot n^{0.2871}$ 

$\implies$ $n \le (\frac{k'+1}{c})^{3.484}$.

The size of $R$ with $M$ of size $k'+1$ is $\mathcal{O}((k'+1)^{3.484})$. This proves that the order of the 6-regular Ramanujan graph with MSMD(3) of size at least $k'+1$ is a polynomial function of $k'$. \vspace{3mm} \qed
\noindent \textbf{Lemma 5.} There is no defensive alliance of size at most $k'$ from the 6-regular Ramanujan graph of size polynomial in $k'$. \vspace{2mm} \\
\textit{Proof.} Let $R$ be the 6-regular Ramanujan graph. Each vertex of $R$ has a degree of six with a closed neighbourhood of seven. To obtain the optimal defensive alliance from $R$, we compute an induced subgraph with each vertex having a closed neighbourhood of four, hence the MSMD(3). We know that the size of MSMD(3) is at least $k'+1$. This concludes that there is no defensive alliance of size at most $k'$ from $R$.\vspace{3mm} \qed
\noindent \textbf{Note:} The construction of Ramanujan graphs by Lubotzky et al. can be done in polynomial time\cite{Ramanujan}. One can construct $r+1$-regular Ramanujan graph using this method when $r$ is a prime and $r \equiv 1 \pmod {4}$. Therefore, the 6-regular Ramanujan graph can be constructed in polynomial time. \vspace{2mm} \\
\indent Let us consider a \textit{6-regular Ramanujan graph with one missing edge} with the same girth as that of a 6-regular Ramanujan graph. We can obtain the graph by removing an edge which is not part of a shortest cycle. In a \textit{6-regular Ramanujan graph with one missing edge}, the vertices of \textit{MSMD(3) with at most one missing edge} can also form a defensive alliance. So, we look to find an \textit{MSMD(3) with at most one missing edge}. The size of \textit{MSMD(3) with at most one missing edge} of girth $g$, is at least the size of MSMD(3) of girth $g$. Therefore, Lemma 4 and Lemma 5 also hold for \textit{6-regular Ramanujan graph with one missing edge}. Hence, we obtain the following lemma. \vspace{2mm} \\
\noindent \textbf{Lemma 6.} There is no defensive alliance of size at most $k'$ from the \textit{6-regular Ramanujan graph with one missing edge }of size polynomial in $k'$. \vspace{2mm} \\
\indent We use the \textit{6-regular Ramanujan graph with one missing edge} as the gadget. The vertices corresponding to the missing edge will be used in place of the forbidden vertices. From Lemma 6, we conclude that there is no defensive alliance of size at most $k'$ from the gadget. \vspace{3mm} \\
\noindent\textbf{Reduction from the D\scriptsize{OMINATING SET ON CUBIC GRAPHS: }\normalsize\vspace{3mm} \\}
We construct an instance $I' = (G', k')$ of the D\scriptsize{EFENSIVE ALLIANCE }\normalsize problem in the following way. See \autoref{fig:Fig2} for an illustration.
\setlength{\textfloatsep}{1\baselineskip plus 0\baselineskip minus 0\baselineskip}
\begin{figure} 
\centering
    \begin{tikzpicture} [scale = 0.6]
    \filldraw[cyan] (0, 0) circle (0.15);
    \filldraw[olive] (2, 1) circle (0.15);
    \filldraw[magenta] (2, -1) circle (0.15);
    \filldraw[blue](4, 1) circle (0.15);
    \filldraw[violet] (4, -1) circle (0.15);
    \filldraw[green] (6, 0) circle (0.15);
    \draw[thin] (0.1, 0.1) -- (1.9, 0.9);
    \draw[thin] (0.1, -0.1) -- (1.9, -0.9);
    \draw[thin] (0.15, 0) -- (5.85, 0);
    \draw[thin] (2, 0.85) -- (2, -0.85);
    \draw[thin] (4, 0.85) -- (4, -0.85);
    \draw[thin] (2.15, 1) -- (3.85, 1);
    \draw[thin] (2.15, -1) -- (3.85, -1);
    \draw[thin] (4.1, 0.9) -- (5.9, 0.1);
    \draw[thin] (4.1, -0.9) -- (5.9, -0.1);
    \filldraw (0, 0.1) circle (0cm) node[anchor=south]{$v_1$};
    \filldraw (2, 1.1) circle (0cm) node[anchor=south]{$v_2$};
    \filldraw (2, -1.8) circle (0cm) node[anchor=south]{$v_3$};
    \filldraw (4, 1.1) circle (0cm) node[anchor=south]{$v_4$};
    \filldraw (4, -1.8) circle (0cm) node[anchor=south]{$v_5$};
    \filldraw (6, 0.1) circle (0cm) node[anchor=south]{$v_6$};
    \filldraw (6, -4) circle (0cm) node[anchor=south]{};
    \filldraw (-2, 0) circle (0cm) node[anchor=south]{};
    \filldraw (8, 0) circle (0cm) node[anchor=south]{$G$};
    \end{tikzpicture}
    \begin{tikzpicture} [scale = 0.62]
    \filldraw (11.5, 6) circle (0cm) node[anchor=south]{$G'$};
        
    \filldraw[green] (0, -4) circle (0.1);
    \filldraw[green] (4, -4) circle (0.1);
    \filldraw[green] (6.5, -4) circle (0.1);
    \filldraw[green] (9,-4) circle (0.1);

    \filldraw[red](-0.2,-4.7) rectangle ++(4pt,4pt);
    \filldraw[red](0,-4.8) rectangle ++(4pt,4pt);

    \draw[thin] (0, -4.1) -- (-0.15,-4.55);
    \draw[thin] (0, -4.1) -- (0.1,-4.65);
    
    \filldraw(3.5, -3) circle (0.1);
    \filldraw(4.5, -3) circle (0.1);
    \filldraw(6, -3) circle (0.1);
    \filldraw(7, -3) circle (0.1);
    \filldraw(8.5, -3) circle (0.1);
    \filldraw(9.5, -3) circle (0.1);
    
    \filldraw(1.25, -4.75) circle (0.1);

    \filldraw[red](4, -4.75) rectangle ++(4pt,4pt);
    \filldraw[red](4.25, -4.75) rectangle ++(4pt,4pt);

    \draw[thin] (4, -4.1) -- (4.05, -4.6);
    \draw[thin] (4, -4.1) -- (4.3, -4.6);

    \filldraw[red](6.5, -4.75) rectangle ++(4pt,4pt);
    \filldraw[red](6.75, -4.75) rectangle ++(4pt,4pt);

    \draw[thin] (6.5, -4.1) -- (6.55, -4.6);
    \draw[thin] (6.5, -4.1) -- (6.8, -4.6);

    \filldraw[red](9, -4.75) rectangle ++(4pt,4pt);
    \filldraw[red](9.25, -4.75) rectangle ++(4pt,4pt);

    \draw[thin] (9, -4.1) -- (9.05, -4.6);
    \draw[thin] (9, -4.1) -- (9.3, -4.6);

    \filldraw[red](3.2, -2.55) rectangle ++(4pt,4pt);
    \filldraw[red](3.5, -2.35) rectangle ++(4pt,4pt);
    \filldraw[red](3.8, -2.55) rectangle ++(4pt,4pt);
    \draw[thin] (3.5, -3) -- (3.3, -2.55);
    \draw[thin] (3.5, -3) -- (3.55, -2.35);
    \draw[thin] (3.5, -3) -- (3.8, -2.55);

    \filldraw[red](4.2, -2.55) rectangle ++(4pt,4pt);
    \filldraw[red](4.5, -2.35) rectangle ++(4pt,4pt);
    \filldraw[red](4.8, -2.55) rectangle ++(4pt,4pt);
    \draw[thin] (4.5, -3) -- (4.3, -2.55);
    \draw[thin] (4.5, -3) -- (4.55, -2.35);
    \draw[thin] (4.5, -3) -- (4.8, -2.55);
    
    \filldraw[red](5.7, -2.55) rectangle ++(4pt,4pt);
    \filldraw[red](6, -2.35) rectangle ++(4pt,4pt);
    \filldraw[red](6.3, -2.55) rectangle ++(4pt,4pt);
    \draw[thin] (6, -3) -- (5.8, -2.55);
    \draw[thin] (6, -3) -- (6.05, -2.35);
    \draw[thin] (6, -3) -- (6.3, -2.55);

    \filldraw[red](6.7, -2.55) rectangle ++(4pt,4pt);
    \filldraw[red](7, -2.35) rectangle ++(4pt,4pt);
    \filldraw[red](7.3, -2.55) rectangle ++(4pt,4pt);
    \draw[thin] (7, -3) -- (6.8, -2.55);
    \draw[thin] (7, -3) -- (7.05, -2.35);
    \draw[thin] (7, -3) -- (7.3, -2.55);

    \filldraw[red](8.2, -2.55) rectangle ++(4pt,4pt);
    \filldraw[red](8.5, -2.35) rectangle ++(4pt,4pt);
    \filldraw[red](8.8, -2.55) rectangle ++(4pt,4pt);
    \draw[thin] (8.5, -3) -- (8.3, -2.55);
    \draw[thin] (8.5, -3) -- (8.55, -2.35);
    \draw[thin] (8.5, -3) -- (8.8, -2.55);

    \filldraw[red](9.2, -2.55) rectangle ++(4pt,4pt);
    \filldraw[red](9.5, -2.35) rectangle ++(4pt,4pt);
    \filldraw[red](9.8, -2.55) rectangle ++(4pt,4pt);
    \draw[thin] (9.5, -3) -- (9.3, -2.55);
    \draw[thin] (9.5, -3) -- (9.55, -2.35);
    \draw[thin] (9.5, -3) -- (9.8, -2.55);

    \draw[thin, black] (0.1, -4) -- (3.5, -3);
    \draw[thin, black] (4, -4) -- (3.5, -3);
    \draw[thin, black] (4, -4) -- (4.5, -3);
    \draw[thin, black] (3.5, -3) -- (4.5, -3);

    \draw[thin, black] (4, -3) -- (6, -3);
    \draw[thin, black] (6.5, -4) -- (6, -3);
    \draw[thin, black] (6.5, -4) -- (7, -3);
    \draw[thin, black] (6, -3) -- (7, -3);

    \draw[thin, black] (6.5, -3) -- (8.5, -3);
    \draw[thin, black] (9, -4) -- (8.5, -3);
    \draw[thin, black] (9, -4) -- (9.5, -3);
    \draw[thin, black] (8.5, -3) -- (9.5, -3);
    
    \filldraw(-1.5, -3) circle (0.1);
    \filldraw(-1.5, -5) circle (0.1);

    \filldraw[red] (-2.35, -5) rectangle ++(4pt,4pt);
    \filldraw[red] (-2.05, -4.7) rectangle ++(4pt,4pt);
    \filldraw[red] (-2.05, -5.3) rectangle ++(4pt,4pt);

    \filldraw[red] (-2.35, -3) rectangle ++(4pt,4pt);
    \filldraw[red] (-2.05, -2.7) rectangle ++(4pt,4pt);
    \filldraw[red] (-2.05, -3.3) rectangle ++(4pt,4pt);
    
    \filldraw[violet] (0, 0) circle (0.1);
    \filldraw[violet] (4, 0) circle (0.1);
    \filldraw[violet] (6.5, 0) circle (0.1);
    \filldraw[violet] (9,0) circle (0.1);

    \filldraw[red](-0.2,-0.7) rectangle ++(4pt,4pt);
    \filldraw[red](0,-0.8) rectangle ++(4pt,4pt);

    \draw[thin] (0, -0.1) -- (-0.15,-0.55);
    \draw[thin] (0, -0.1) -- (0.1,-0.65);
    
    \filldraw(3.5, 1) circle (0.1);
    \filldraw(4.5, 1) circle (0.1);
    \filldraw(6, 1) circle (0.1);
    \filldraw(7, 1) circle (0.1);
    \filldraw(8.5, 1) circle (0.1);
    \filldraw(9.5, 1) circle (0.1);

    \filldraw(1.25, -0.75) circle (0.1);

    \filldraw[red](4, -0.75) rectangle ++(4pt,4pt);
    \filldraw[red](4.25, -0.75) rectangle ++(4pt,4pt);

    \draw[thin] (4, -0.1) -- (4.05, -0.6);
    \draw[thin] (4, -0.1) -- (4.3, -0.6);

    \filldraw[red](6.5, -0.75) rectangle ++(4pt,4pt);
    \filldraw[red](6.75, -0.75) rectangle ++(4pt,4pt);

    \draw[thin] (6.5, -0.1) -- (6.55, -0.6);
    \draw[thin] (6.5, -0.1) -- (6.8, -0.6);

    \filldraw[red](9, -0.75) rectangle ++(4pt,4pt);
    \filldraw[red](9.25, -0.75) rectangle ++(4pt,4pt);

    \draw[thin] (9, -0.1) -- (9.05, -0.6);
    \draw[thin] (9, -0.1) -- (9.3, -0.6);

    \filldraw[red](3.2, 1.45) rectangle ++(4pt,4pt);
    \filldraw[red](3.5, 1.65) rectangle ++(4pt,4pt);
    \filldraw[red](3.8, 1.45) rectangle ++(4pt,4pt);
    \draw[thin] (3.5, 1) -- (3.3, 1.45);
    \draw[thin] (3.5, 1) -- (3.55, 1.65);
    \draw[thin] (3.5, 1) -- (3.8, 1.45);

    \filldraw[red](4.2, 1.45) rectangle ++(4pt,4pt);
    \filldraw[red](4.5, 1.65) rectangle ++(4pt,4pt);
    \filldraw[red](4.8, 1.45) rectangle ++(4pt,4pt);
    \draw[thin] (4.5, 1) -- (4.3, 1.45);
    \draw[thin] (4.5, 1) -- (4.55, 1.65);
    \draw[thin] (4.5, 1) -- (4.8, 1.45);
    
    \filldraw[red](5.7, 1.45) rectangle ++(4pt,4pt);
    \filldraw[red](6, 1.65) rectangle ++(4pt,4pt);
    \filldraw[red](6.3, 1.45) rectangle ++(4pt,4pt);
    \draw[thin] (6, 1) -- (5.8, 1.45);
    \draw[thin] (6, 1) -- (6.05, 1.65);
    \draw[thin] (6, 1) -- (6.3, 1.45);

    \filldraw[red](6.7, 1.45) rectangle ++(4pt,4pt);
    \filldraw[red](7, 1.65) rectangle ++(4pt,4pt);
    \filldraw[red](7.3, 1.45) rectangle ++(4pt,4pt);
    \draw[thin] (7, 1) -- (6.8, 1.45);
    \draw[thin] (7, 1) -- (7.05, 1.65);
    \draw[thin] (7, 1) -- (7.3, 1.45);

    \filldraw[red](8.2, 1.45) rectangle ++(4pt,4pt);
    \filldraw[red](8.5, 1.65) rectangle ++(4pt,4pt);
    \filldraw[red](8.8, 1.45) rectangle ++(4pt,4pt);
    \draw[thin] (8.5, 1) -- (8.3, 1.45);
    \draw[thin] (8.5, 1) -- (8.55, 1.65);
    \draw[thin] (8.5, 1) -- (8.8, 1.45);

    \filldraw[red](9.2, 1.45) rectangle ++(4pt,4pt);
    \filldraw[red](9.5, 1.65) rectangle ++(4pt,4pt);
    \filldraw[red](9.8, 1.45) rectangle ++(4pt,4pt);
    \draw[thin] (9.5, 1) -- (9.3, 1.45);
    \draw[thin] (9.5, 1) -- (9.55, 1.65);
    \draw[thin] (9.5, 1) -- (9.8, 1.45);

    \draw[thin, black] (0.1, 0) -- (3.5, 1);
    \draw[thin, black] (4, 0) -- (3.5, 1);
    \draw[thin, black] (4, 0) -- (4.5, 1);
    \draw[thin, black] (3.5, 1) -- (4.5, 1);

    \draw[thin, black] (4, 1) -- (6, 1);
    \draw[thin, black] (6.5, 0) -- (6, 1);
    \draw[thin, black] (6.5, 0) -- (7, 1);
    \draw[thin, black] (6, 1) -- (7, 1);

    \draw[thin, black] (6.5, 1) -- (8.5, 1);
    \draw[thin, black] (9, 0) -- (8.5, 1);
    \draw[thin, black] (9, 0) -- (9.5, 1);
    \draw[thin, black] (8.5, 1) -- (9.5, 1);
    
    \filldraw(-1.5, 1) circle (0.1);
    \filldraw(-1.5, -1) circle (0.1);

    \filldraw[red] (-2.35, -1) rectangle ++(4pt,4pt);
    \filldraw[red] (-2.05, -0.7) rectangle ++(4pt,4pt);
    \filldraw[red] (-2.05, -1.3) rectangle ++(4pt,4pt);

    \filldraw[red] (-2.35, 1) rectangle ++(4pt,4pt);
    \filldraw[red] (-2.05, 1.3) rectangle ++(4pt,4pt);
    \filldraw[red] (-2.05, 0.7) rectangle ++(4pt,4pt);  

    \filldraw[blue] (0, 4) circle (0.1);
    \filldraw[blue] (4, 4) circle (0.1);
    \filldraw[blue] (6.5, 4) circle (0.1);
    \filldraw[blue] (9, 4) circle (0.1);

    \filldraw(3.5, 5) circle (0.1);
    \filldraw(4.5, 5) circle (0.1);
    \filldraw(6, 5) circle (0.1);
    \filldraw(7, 5) circle (0.1);
    \filldraw(8.5, 5) circle (0.1);
    \filldraw(9.5, 5) circle (0.1);
    
    \filldraw(1.25, 3.25) circle (0.1);

    \filldraw[red](-0.2, 3.3) rectangle ++(4pt,4pt);
    \filldraw[red](0,3.2) rectangle ++(4pt,4pt);

    \draw[thin] (0, 3.9) -- (-0.15,3.45);
    \draw[thin] (0, 3.9) -- (0.1,3.35);

    \filldraw[red](4, 3.25) rectangle ++(4pt,4pt);
    \filldraw[red](4.25, 3.25) rectangle ++(4pt,4pt);

    \draw[thin] (4, 3.9) -- (4.05, 3.4);
    \draw[thin] (4, 3.9) -- (4.3, 3.4);

    \filldraw[red](6.5, 3.25) rectangle ++(4pt,4pt);
    \filldraw[red](6.75, 3.25) rectangle ++(4pt,4pt);

    \draw[thin] (6.5, 3.9) -- (6.55, 3.4);
    \draw[thin] (6.5, 3.9) -- (6.8, 3.4);

    \filldraw[red](9, 3.25) rectangle ++(4pt,4pt);
    \filldraw[red](9.25, 3.25) rectangle ++(4pt,4pt);

    \draw[thin] (9, 3.9) -- (9.05, 3.4);
    \draw[thin] (9, 3.9) -- (9.3, 3.4);

    \filldraw[red](3.2, 5.45) rectangle ++(4pt,4pt);
    \filldraw[red](3.5, 5.65) rectangle ++(4pt,4pt);
    \filldraw[red](3.8, 5.45) rectangle ++(4pt,4pt);
    \draw[thin] (3.5, 5) -- (3.3, 5.45);
    \draw[thin] (3.5, 5) -- (3.55, 5.65);
    \draw[thin] (3.5, 5) -- (3.8, 5.45);

    \filldraw[red](4.2, 5.45) rectangle ++(4pt,4pt);
    \filldraw[red](4.5, 5.65) rectangle ++(4pt,4pt);
    \filldraw[red](4.8, 5.45) rectangle ++(4pt,4pt);
    \draw[thin] (4.5, 5) -- (4.3, 5.45);
    \draw[thin] (4.5, 5) -- (4.55, 5.65);
    \draw[thin] (4.5, 5) -- (4.8, 5.45);
    
    \filldraw[red](5.7, 5.45) rectangle ++(4pt,4pt);
    \filldraw[red](6, 5.65) rectangle ++(4pt,4pt);
    \filldraw[red](6.3, 5.45) rectangle ++(4pt,4pt);
    \draw[thin] (6, 5) -- (5.8, 5.45);
    \draw[thin] (6, 5) -- (6.05, 5.65);
    \draw[thin] (6, 5) -- (6.3, 5.45);

    \filldraw[red](6.7, 5.45) rectangle ++(4pt,4pt);
    \filldraw[red](7, 5.65) rectangle ++(4pt,4pt);
    \filldraw[red](7.3, 5.45) rectangle ++(4pt,4pt);
    \draw[thin] (7, 5) -- (6.8, 5.45);
    \draw[thin] (7, 5) -- (7.05, 5.65);
    \draw[thin] (7, 5) -- (7.3, 5.45);

    \filldraw[red](8.2, 5.45) rectangle ++(4pt,4pt);
    \filldraw[red](8.5, 5.65) rectangle ++(4pt,4pt);
    \filldraw[red](8.8, 5.45) rectangle ++(4pt,4pt);
    \draw[thin] (8.5, 5) -- (8.3, 5.45);
    \draw[thin] (8.5, 5) -- (8.55, 5.65);
    \draw[thin] (8.5, 5) -- (8.8, 5.45);

    \filldraw[red](9.2, 5.45) rectangle ++(4pt,4pt);
    \filldraw[red](9.5, 5.65) rectangle ++(4pt,4pt);
    \filldraw[red](9.8, 5.45) rectangle ++(4pt,4pt);
    \draw[thin] (9.5, 5) -- (9.3, 5.45);
    \draw[thin] (9.5, 5) -- (9.55, 5.65);
    \draw[thin] (9.5, 5) -- (9.8, 5.45);

    \draw[thin, black] (0.1, 4) -- (3.5, 5);
    \draw[thin, black] (4, 4) -- (3.5, 5);
    \draw[thin, black] (4, 4) -- (4.5, 5);
    \draw[thin, black] (3.5, 5) -- (4.5, 5);
    
    \draw[thin, black] (4, 5) -- (6, 5);
    \draw[thin, black] (6.5, 4) -- (6, 5);
    \draw[thin, black] (6.5, 4) -- (7, 5);
    \draw[thin, black] (6, 5) -- (7, 5);

    \draw[thin, black] (6.5, 5) -- (8.5, 5);
    \draw[thin, black] (9, 4) -- (8.5, 5);
    \draw[thin, black] (9, 4) -- (9.5, 5);
    \draw[thin, black] (8.5, 5) -- (9.5, 5);
    
    \filldraw(-1.5, 5) circle (0.1);
    \filldraw(-1.5, 3) circle (0.1);

    \filldraw[red] (-2.35, 3) rectangle ++(4pt,4pt);
    \filldraw[red] (-2.05, 3.3) rectangle ++(4pt,4pt);
    \filldraw[red] (-2.05, 2.7) rectangle ++(4pt,4pt);

    \filldraw[red] (-2.35, 5) rectangle ++(4pt,4pt);
    \filldraw[red] (-2.05, 5.3) rectangle ++(4pt,4pt);
    \filldraw[red] (-2.05, 4.7) rectangle ++(4pt,4pt);     

    \filldraw[magenta] (0, 8) circle (0.1);
    \filldraw[magenta] (4, 8) circle (0.1);
    \filldraw[magenta] (6.5, 8) circle (0.1);
    \filldraw[magenta] (9, 8) circle (0.1);

    \filldraw(3.5, 9) circle (0.1);
    \filldraw(4.5, 9) circle (0.1);
    \filldraw(6, 9) circle (0.1);
    \filldraw(7, 9) circle (0.1);
    \filldraw(8.5, 9) circle (0.1);
    \filldraw(9.5, 9) circle (0.1);

    \filldraw(1.25, 7.25) circle (0.1);

    \filldraw[red](-0.2, 7.3) rectangle ++(4pt,4pt);
    \filldraw[red](0,7.2) rectangle ++(4pt,4pt);

    \draw[thin] (0, 7.9) -- (-0.15,7.45);
    \draw[thin] (0, 7.9) -- (0.1,7.35);

    \filldraw[red](4, 7.25) rectangle ++(4pt,4pt);
    \filldraw[red](4.25, 7.25) rectangle ++(4pt,4pt);

    \draw[thin] (4, 7.9) -- (4.05, 7.4);
    \draw[thin] (4, 7.9) -- (4.3, 7.4);

    \filldraw[red](6.5, 7.25) rectangle ++(4pt,4pt);
    \filldraw[red](6.75, 7.25) rectangle ++(4pt,4pt);

    \draw[thin] (6.5, 7.9) -- (6.55, 7.4);
    \draw[thin] (6.5, 7.9) -- (6.8, 7.4);

    \filldraw[red](9, 7.25) rectangle ++(4pt,4pt);
    \filldraw[red](9.25, 7.25) rectangle ++(4pt,4pt);

    \draw[thin] (9, 7.9) -- (9.05, 7.4);
    \draw[thin] (9, 7.9) -- (9.3, 7.4);

    \filldraw[red](3.2, 9.45) rectangle ++(4pt,4pt);
    \filldraw[red](3.5, 9.65) rectangle ++(4pt,4pt);
    \filldraw[red](3.8, 9.45) rectangle ++(4pt,4pt);
    \draw[thin] (3.5, 9) -- (3.3, 9.45);
    \draw[thin] (3.5, 9) -- (3.55, 9.65);
    \draw[thin] (3.5, 9) -- (3.8, 9.45);

    \filldraw[red](4.2, 9.45) rectangle ++(4pt,4pt);
    \filldraw[red](4.5, 9.65) rectangle ++(4pt,4pt);
    \filldraw[red](4.8, 9.45) rectangle ++(4pt,4pt);
    \draw[thin] (4.5, 9) -- (4.3, 9.45);
    \draw[thin] (4.5, 9) -- (4.55, 9.65);
    \draw[thin] (4.5, 9) -- (4.8, 9.45);
    
    \filldraw[red](5.7, 9.45) rectangle ++(4pt,4pt);
    \filldraw[red](6, 9.65) rectangle ++(4pt,4pt);
    \filldraw[red](6.3, 9.45) rectangle ++(4pt,4pt);
    \draw[thin] (6, 9) -- (5.8, 9.45);
    \draw[thin] (6, 9) -- (6.05, 9.65);
    \draw[thin] (6, 9) -- (6.3, 9.45);

    \filldraw[red](6.7, 9.45) rectangle ++(4pt,4pt);
    \filldraw[red](7, 9.65) rectangle ++(4pt,4pt);
    \filldraw[red](7.3, 9.45) rectangle ++(4pt,4pt);
    \draw[thin] (7, 9) -- (6.8, 9.45);
    \draw[thin] (7, 9) -- (7.05, 9.65);
    \draw[thin] (7, 9) -- (7.3, 9.45);

    \filldraw[red](8.2, 9.45) rectangle ++(4pt,4pt);
    \filldraw[red](8.5, 9.65) rectangle ++(4pt,4pt);
    \filldraw[red](8.8, 9.45) rectangle ++(4pt,4pt);
    \draw[thin] (8.5, 9) -- (8.3, 9.45);
    \draw[thin] (8.5, 9) -- (8.55, 9.65);
    \draw[thin] (8.5, 9) -- (8.8, 9.45);

    \filldraw[red](9.2, 9.45) rectangle ++(4pt,4pt);
    \filldraw[red](9.5, 9.65) rectangle ++(4pt,4pt);
    \filldraw[red](9.8, 9.45) rectangle ++(4pt,4pt);
    \draw[thin] (9.5, 9) -- (9.3, 9.45);
    \draw[thin] (9.5, 9) -- (9.55, 9.65);
    \draw[thin] (9.5, 9) -- (9.8, 9.45);

    \draw[thin, black] (0.1, 8) -- (3.5, 9);
    \draw[thin, black] (4, 8) -- (3.5, 9);
    \draw[thin, black] (4, 8) -- (4.5, 9);
    \draw[thin, black] (3.5, 9) -- (4.5, 9);
    
    \draw[thin, black] (4, 9) -- (6, 9);
    \draw[thin, black] (6.5, 8) -- (6, 9);
    \draw[thin, black] (6.5, 8) -- (7, 9);
    \draw[thin, black] (6, 9) -- (7, 9);

    \draw[thin, black] (6.5, 9) -- (8.5, 9);
    \draw[thin, black] (9, 8) -- (8.5, 9);
    \draw[thin, black] (9, 8) -- (9.5, 9);
    \draw[thin, black] (8.5, 9) -- (9.5, 9); 
    
     \filldraw(-1.5, 9) circle (0.1);
    \filldraw(-1.5, 7) circle (0.1);

    \filldraw[red] (-2.35, 7) rectangle ++(4pt,4pt);
    \filldraw[red] (-2.05, 7.3) rectangle ++(4pt,4pt);
    \filldraw[red] (-2.05, 6.7) rectangle ++(4pt,4pt);

    \filldraw[red] (-2.35, 9) rectangle ++(4pt,4pt);
    \filldraw[red] (-2.05, 9.3) rectangle ++(4pt,4pt);
    \filldraw[red] (-2.05, 8.7) rectangle ++(4pt,4pt);   

    \filldraw[olive] (0, 12) circle (0.1);
    \filldraw[olive] (4, 12) circle (0.1);
    \filldraw[olive] (6.5, 12) circle (0.1);
    \filldraw[olive] (9, 12) circle (0.1);

    \filldraw(3.5, 13) circle (0.1);
    \filldraw(4.5, 13) circle (0.1);
    \filldraw(6, 13) circle (0.1);
    \filldraw(7, 13) circle (0.1);
    \filldraw(8.5, 13) circle (0.1);
    \filldraw(9.5, 13) circle (0.1);
    
    \filldraw(1.25, 11.25) circle (0.1);

    \filldraw[red](-0.2, 11.3) rectangle ++(4pt,4pt);
    \filldraw[red](0,11.2) rectangle ++(4pt,4pt);

    \draw[thin] (0, 11.9) -- (-0.15,11.45);
    \draw[thin] (0, 11.9) -- (0.1,11.35);

    \filldraw[red](4, 11.25) rectangle ++(4pt,4pt);
    \filldraw[red](4.25, 11.25) rectangle ++(4pt,4pt);

    \draw[thin] (4, 11.9) -- (4.05, 11.4);
    \draw[thin] (4, 11.9) -- (4.3, 11.4);

    \filldraw[red](6.5, 11.25) rectangle ++(4pt,4pt);
    \filldraw[red](6.75, 11.25) rectangle ++(4pt,4pt);

    \draw[thin] (6.5, 11.9) -- (6.55, 11.4);
    \draw[thin] (6.5, 11.9) -- (6.8, 11.4);

    \filldraw[red](9, 11.25) rectangle ++(4pt,4pt);
    \filldraw[red](9.25, 11.25) rectangle ++(4pt,4pt);

    \draw[thin] (9, 11.9) -- (9.05, 11.4);
    \draw[thin] (9, 11.9) -- (9.3, 11.4);

    \filldraw[red](3.2, 13.45) rectangle ++(4pt,4pt);
    \filldraw[red](3.5, 13.65) rectangle ++(4pt,4pt);
    \filldraw[red](3.8, 13.45) rectangle ++(4pt,4pt);
    \draw[thin] (3.5, 13) -- (3.3, 13.45);
    \draw[thin] (3.5, 13) -- (3.55, 13.65);
    \draw[thin] (3.5, 13) -- (3.8, 13.45);

    \filldraw[red](4.2, 13.45) rectangle ++(4pt,4pt);
    \filldraw[red](4.5, 13.65) rectangle ++(4pt,4pt);
    \filldraw[red](4.8, 13.45) rectangle ++(4pt,4pt);
    \draw[thin] (4.5, 13) -- (4.3, 13.45);
    \draw[thin] (4.5, 13) -- (4.55, 13.65);
    \draw[thin] (4.5, 13) -- (4.8, 13.45);
    
    \filldraw[red](5.7, 13.45) rectangle ++(4pt,4pt);
    \filldraw[red](6, 13.65) rectangle ++(4pt,4pt);
    \filldraw[red](6.3, 13.45) rectangle ++(4pt,4pt);
    \draw[thin] (6, 13) -- (5.8, 13.45);
    \draw[thin] (6, 13) -- (6.05, 13.65);
    \draw[thin] (6, 13) -- (6.3, 13.45);

    \filldraw[red](6.7, 13.45) rectangle ++(4pt,4pt);
    \filldraw[red](7, 13.65) rectangle ++(4pt,4pt);
    \filldraw[red](7.3, 13.45) rectangle ++(4pt,4pt);
    \draw[thin] (7, 13) -- (6.8, 13.45);
    \draw[thin] (7, 13) -- (7.05, 13.65);
    \draw[thin] (7, 13) -- (7.3, 13.45);

    \filldraw[red](8.2, 13.45) rectangle ++(4pt,4pt);
    \filldraw[red](8.5, 13.65) rectangle ++(4pt,4pt);
    \filldraw[red](8.8, 13.45) rectangle ++(4pt,4pt);
    \draw[thin] (8.5, 13) -- (8.3, 13.45);
    \draw[thin] (8.5, 13) -- (8.55, 13.65);
    \draw[thin] (8.5, 13) -- (8.8, 13.45);

    \filldraw[red](9.2, 13.45) rectangle ++(4pt,4pt);
    \filldraw[red](9.5, 13.65) rectangle ++(4pt,4pt);
    \filldraw[red](9.8, 13.45) rectangle ++(4pt,4pt);
    \draw[thin] (9.5, 13) -- (9.3, 13.45);
    \draw[thin] (9.5, 13) -- (9.55, 13.65);
    \draw[thin] (9.5, 13) -- (9.8, 13.45);

    \draw[thin, black] (0.1, 12) -- (3.5, 13);
    \draw[thin, black] (4, 12) -- (3.5, 13);
    \draw[thin, black] (4, 12) -- (4.5, 13);
    \draw[thin, black] (3.5, 13) -- (4.5, 13);
    
    \draw[thin, black] (4, 13) -- (6, 13);
    \draw[thin, black] (6.5, 12) -- (6, 13);
    \draw[thin, black] (6.5, 12) -- (7, 13);
    \draw[thin, black] (6, 13) -- (7, 13);

    \draw[thin, black] (6.5, 13) -- (8.5, 13);
    \draw[thin, black] (9, 12) -- (8.5, 13);
    \draw[thin, black] (9, 12) -- (9.5, 13);
    \draw[thin, black] (8.5, 13) -- (9.5, 13); 

    \filldraw(-1.5, 13) circle (0.1);
    \filldraw(-1.5, 11) circle (0.1);

    \filldraw[red] (-2.35, 11) rectangle ++(4pt,4pt);
    \filldraw[red] (-2.05, 11.3) rectangle ++(4pt,4pt);
    \filldraw[red] (-2.05, 10.7) rectangle ++(4pt,4pt);

    \filldraw[red] (-2.35, 13) rectangle ++(4pt,4pt);
    \filldraw[red] (-2.05, 13.3) rectangle ++(4pt,4pt);
    \filldraw[red] (-2.05, 12.7) rectangle ++(4pt,4pt);   

    \filldraw[cyan] (0, 16) circle (0.1);
    \filldraw[cyan] (4, 16) circle (0.1);
    \filldraw[cyan] (6.5, 16) circle (0.1);
    \filldraw[cyan] (9, 16) circle (0.1);

    \filldraw(3.5, 17) circle (0.1);
    \filldraw(4.5, 17) circle (0.1);
    \filldraw(6, 17) circle (0.1);
    \filldraw(7, 17) circle (0.1);
    \filldraw(8.5, 17) circle (0.1);
    \filldraw(9.5, 17) circle (0.1);

    \filldraw(1.25, 15.25) circle (0.1);

    \filldraw[red](-0.2, 15.3) rectangle ++(4pt,4pt);
    \filldraw[red](0,15.2) rectangle ++(4pt,4pt);

    \draw[thin] (0, 15.9) -- (-0.15,15.45);
    \draw[thin] (0, 15.9) -- (0.1,15.35);

    \filldraw[red](4, 15.25) rectangle ++(4pt,4pt);
    \filldraw[red](4.25, 15.25) rectangle ++(4pt,4pt);

    \draw[thin] (4, 15.9) -- (4.05, 15.4);
    \draw[thin] (4, 15.9) -- (4.3, 15.4);

    \filldraw[red](6.5, 15.25) rectangle ++(4pt,4pt);
    \filldraw[red](6.75, 15.25) rectangle ++(4pt,4pt);

    \draw[thin] (6.5, 15.9) -- (6.55, 15.4);
    \draw[thin] (6.5, 15.9) -- (6.8, 15.4);

    \filldraw[red](9, 15.25) rectangle ++(4pt,4pt);
    \filldraw[red](9.25, 15.25) rectangle ++(4pt,4pt);

    \draw[thin] (9, 15.9) -- (9.05, 15.4);
    \draw[thin] (9, 15.9) -- (9.3, 15.4);

    \filldraw[red](3.2, 17.45) rectangle ++(4pt,4pt);
    \filldraw[red](3.5, 17.65) rectangle ++(4pt,4pt);
    \filldraw[red](3.8, 17.45) rectangle ++(4pt,4pt);
    \draw[thin] (3.5, 17) -- (3.3, 17.45);
    \draw[thin] (3.5, 17) -- (3.55, 17.65);
    \draw[thin] (3.5, 17) -- (3.8, 17.45);

    \filldraw[red](4.2, 17.45) rectangle ++(4pt,4pt);
    \filldraw[red](4.5, 17.65) rectangle ++(4pt,4pt);
    \filldraw[red](4.8, 17.45) rectangle ++(4pt,4pt);
    \draw[thin] (4.5, 17) -- (4.3, 17.45);
    \draw[thin] (4.5, 17) -- (4.55, 17.65);
    \draw[thin] (4.5, 17) -- (4.8, 17.45);
    
    \filldraw[red](5.7, 17.45) rectangle ++(4pt,4pt);
    \filldraw[red](6, 17.65) rectangle ++(4pt,4pt);
    \filldraw[red](6.3, 17.45) rectangle ++(4pt,4pt);
    \draw[thin] (6, 17) -- (5.8, 17.45);
    \draw[thin] (6, 17) -- (6.05, 17.65);
    \draw[thin] (6, 17) -- (6.3, 17.45);

    \filldraw[red](6.7, 17.45) rectangle ++(4pt,4pt);
    \filldraw[red](7, 17.65) rectangle ++(4pt,4pt);
    \filldraw[red](7.3, 17.45) rectangle ++(4pt,4pt);
    \draw[thin] (7, 17) -- (6.8, 17.45);
    \draw[thin] (7, 17) -- (7.05, 17.65);
    \draw[thin] (7, 17) -- (7.3, 17.45);

    \filldraw[red](8.2, 17.45) rectangle ++(4pt,4pt);
    \filldraw[red](8.5, 17.65) rectangle ++(4pt,4pt);
    \filldraw[red](8.8, 17.45) rectangle ++(4pt,4pt);
    \draw[thin] (8.5, 17) -- (8.3, 17.45);
    \draw[thin] (8.5, 17) -- (8.55, 17.65);
    \draw[thin] (8.5, 17) -- (8.8, 17.45);

    \filldraw[red](9.2, 17.45) rectangle ++(4pt,4pt);
    \filldraw[red](9.5, 17.65) rectangle ++(4pt,4pt);
    \filldraw[red](9.8, 17.45) rectangle ++(4pt,4pt);
    \draw[thin] (9.5, 17) -- (9.3, 17.45);
    \draw[thin] (9.5, 17) -- (9.55, 17.65);
    \draw[thin] (9.5, 17) -- (9.8, 17.45);

    \draw[thin, black] (0.1, 16) -- (3.5, 17);
    \draw[thin, black] (4, 16) -- (3.5, 17);
    \draw[thin, black] (4, 16) -- (4.5, 17);
    \draw[thin, black] (3.5, 17) -- (4.5, 17);
    
    \draw[thin, black] (4, 17) -- (6, 17);
    \draw[thin, black] (6.5, 16) -- (6, 17);
    \draw[thin, black] (6.5, 16) -- (7, 17);
    \draw[thin, black] (6, 17) -- (7, 17);

    \draw[thin, black] (6.5, 17) -- (8.5, 17);
    \draw[thin, black] (9, 16) -- (8.5, 17);
    \draw[thin, black] (9, 16) -- (9.5, 17);
    \draw[thin, black] (8.5, 17) -- (9.5, 17);

    \filldraw(-1.5, 17) circle (0.1);
    \filldraw(-1.5, 15) circle (0.1);

    \filldraw[red] (-2.35, 15) rectangle ++(4pt,4pt);
    \filldraw[red] (-2.05, 15.3) rectangle ++(4pt,4pt);
    \filldraw[red] (-2.05, 14.7) rectangle ++(4pt,4pt);

    \filldraw[red] (-2.35, 17) rectangle ++(4pt,4pt);
    \filldraw[red] (-2.05, 17.3) rectangle ++(4pt,4pt);
    \filldraw[red] (-2.05, 16.7) rectangle ++(4pt,4pt);   

    \draw[thin] (1.25, 15.25) -- (0.1, 16);
    \draw[thin, cyan] (1.25, 15.25) -- (4, 12);
    \draw[thin, cyan] (1.25, 15.25) -- (4, 8);
    \draw[thin, cyan] (1.25, 15.25) -- (4, -4);

    \draw[thin] (1.25, 11.25) -- (0.1, 12);
    \draw[thin, olive] (1.25, 11.25) -- (4, 16);
    \draw[thin, olive] (1.25, 11.25) -- (6.5, 8);
    \draw[thin, olive] (1.25, 11.25) -- (4, 4);

    \draw[thin] (1.25, 7.25) -- (0.1, 8);
    \draw[thin, magenta] (1.25, 7.25) -- (6.5, 16);
    \draw[thin, magenta] (1.25, 7.25) -- (6.5, 12);
    \draw[thin, magenta] (1.25, 7.25) -- (4, 0);

    \draw[thin] (1.25, 3.25) -- (0.1, 4);
    \draw[thin, blue] (1.25, 3.25) -- (9, 12);
    \draw[thin, blue] (1.25, 3.25) -- (6.5, 0);
    \draw[thin, blue] (1.25, 3.25) -- (6.5, -4);

    \draw[thin] (1.25, -0.75) -- (0.1, 0);
    \draw[thin, violet] (1.25, -0.75) -- (9, 8);
    \draw[thin, violet] (1.25, -0.75) -- (6.5, 4);
    \draw[thin, violet] (1.25, -0.75) -- (9, -4);

    \draw[thin] (1.25, -4.75) -- (0.1, -4);
    \draw[thin, green] (1.25, -4.75) -- (9, 16);
    \draw[thin, green] (1.25, -4.75) -- (9, 4);
    \draw[thin, green] (1.25, -4.75) -- (9, 0);
    
    \draw[thin, black] (-0.1, -4) -- (-1.5, -5);
    \draw[thin, black] (-0.1, -4) -- (-1.5, -3);
    \draw[thin, black] (-1.5, -5) -- (-1.5, -3);
    \draw[thin, black] (-1.5, -5) -- (-2.35, -5);
    \draw[thin, black] (-1.5, -5) -- (-2.05, -4.7);
    \draw[thin, black] (-1.5, -5) -- (-2.05, -5.3);
    \draw[thin, black] (-1.5, -3) -- (-2.35, -3);
    \draw[thin, black] (-1.5, -3) -- (-2.05, -2.7);
    \draw[thin, black] (-1.5, -3) -- (-2.05, -3.3);

    \draw[thin, black] (-0.1, 0) -- (-1.5, 1);
    \draw[thin, black] (-0.1, 0) -- (-1.5, -1);
    \draw[thin, black] (-1.5, 1) -- (-1.5, -1);
    \draw[thin] (-1.5, -1) -- (-1.5, -3);
    \draw[thin, black] (-1.5, -1) -- (-2.35, -1);
    \draw[thin, black] (-1.5, -1) -- (-2.05, -0.7);
    \draw[thin, black] (-1.5, -1) -- (-2.05, -1.3);
    \draw[thin, black] (-1.5, 1) -- (-2.35, 1);
    \draw[thin, black] (-1.5, 1) -- (-2.05, 1.3);
    \draw[thin, black] (-1.5, 1) -- (-2.05, 0.7);

    \draw[thin, black] (-0.1, 4) -- (-1.5, 3);
    \draw[thin, black] (-0.1, 4) -- (-1.5, 5);
    \draw[thin, black] (-1.5, 3) -- (-1.5, 5);
    \draw[thin] (-1.5, 1) -- (-1.5, 3);
    \draw[thin, black] (-1.5, 3) -- (-2.35, 3);
    \draw[thin, black] (-1.5, 3) -- (-2.05, 3.3);
    \draw[thin, black] (-1.5, 3) -- (-2.05, 2.7);
    \draw[thin, black] (-1.5, 5) -- (-2.35, 5);
    \draw[thin, black] (-1.5, 5) -- (-2.05, 5.3);
    \draw[thin, black] (-1.5, 5) -- (-2.05, 4.7);

    \draw[thin, black] (-0.1, 8) -- (-1.5, 7);
    \draw[thin, black] (-0.1, 8) -- (-1.5, 9);
    \draw[thin, black] (-1.5, 7) -- (-1.5, 9);
    \draw[thin] (-1.5, 5) -- (-1.5, 7);
    \draw[thin, black] (-1.5, 7) -- (-2.35, 7);
    \draw[thin, black] (-1.5, 7) -- (-2.05, 7.3);
    \draw[thin, black] (-1.5, 7) -- (-2.05, 6.7);
    \draw[thin, black] (-1.5, 9) -- (-2.35, 9);
    \draw[thin, black] (-1.5, 9) -- (-2.05, 9.3);
    \draw[thin, black] (-1.5, 9) -- (-2.05, 8.7);
        
    \draw[thin, black] (-0.1, 12) -- (-1.5, 11);
    \draw[thin, black] (-0.1, 12) -- (-1.5, 13);
    \draw[thin, black] (-1.5, 11) -- (-1.5, 13);
    \draw[thin] (-1.5, 9) -- (-1.5, 11);
    \draw[thin, black] (-1.5, 11) -- (-2.35, 11);
    \draw[thin, black] (-1.5, 11) -- (-2.05, 11.3);
    \draw[thin, black] (-1.5, 11) -- (-2.05, 10.7);
    \draw[thin, black] (-1.5, 13) -- (-2.35, 13);
    \draw[thin, black] (-1.5, 13) -- (-2.05, 13.3);
    \draw[thin, black] (-1.5, 13) -- (-2.05, 12.7);
        
    \draw[thin, black] (-0.1, 16) -- (-1.5, 15);
    \draw[thin, black] (-0.1, 16) -- (-1.5, 17);
    \draw[thin, black] (-1.5, 15) -- (-1.5, 17);
    \draw[thin] (-1.5, 13) -- (-1.5, 15);
    \draw[thin, black] (-1.5, 15) -- (-2.35, 15);
    \draw[thin, black] (-1.5, 15) -- (-2.05, 15.3);
    \draw[thin, black] (-1.5, 15) -- (-2.05, 14.7);
    \draw[thin, black] (-1.5, 17) -- (-2.35, 17);
    \draw[thin, black] (-1.5, 17) -- (-2.05, 17.3);
    \draw[thin, black] (-1.5, 17) -- (-2.05, 16.7);

    \filldraw (0.75, 15.5) circle (0cm) node[anchor=south]{$v_1^0$};
    \filldraw (4.5, 15.5) circle (0cm) node[anchor=south]{$v_1^1$};
    \filldraw (3.2, 16.2) circle (0cm) node[anchor=south]{$u_1^1$};
    \filldraw (4.8, 16.2) circle (0cm) node[anchor=south]{$w_1^1$};
    \filldraw (5.7, 16.2) circle (0cm) node[anchor=south]{$u_1^2$};
    \filldraw (7.3, 16.2) circle (0cm) node[anchor=south]{$w_1^2$};
    \filldraw (8.2, 16.2) circle (0cm) node[anchor=south]{$u_1^3$};
    \filldraw (9.8, 16.2) circle (0cm) node[anchor=south]{$w_1^3$};
    \filldraw (7, 15.5) circle (0cm) node[anchor=south]{$v_1^2$};
    \filldraw (9.5, 15.5) circle (0cm) node[anchor=south]{$v_1^3$};
    
    \filldraw (0.75, 11.5) circle (0cm) node[anchor=south]{$v_2^0$};
    \filldraw (4.5, 11.5) circle (0cm) node[anchor=south]{$v_2^1$};
    \filldraw (3.2, 12.2) circle (0cm) node[anchor=south]{$u_2^1$};
    \filldraw (4.8, 12.2) circle (0cm) node[anchor=south]{$w_2^1$};
    \filldraw (5.7, 12.2) circle (0cm) node[anchor=south]{$u_2^2$};
    \filldraw (7.3, 12.2) circle (0cm) node[anchor=south]{$w_2^2$};
    \filldraw (8.2, 12.2) circle (0cm) node[anchor=south]{$u_2^3$};
    \filldraw (9.8, 12.2) circle (0cm) node[anchor=south]{$w_2^3$};
    \filldraw (7, 11.5) circle (0cm) node[anchor=south]{$v_2^2$};
    \filldraw (9.5, 11.5) circle (0cm) node[anchor=south]{$v_2^3$};
    
    \filldraw (0.75, 7.5) circle (0cm) node[anchor=south]{$v_3^0$};
    \filldraw (4.5, 7.5) circle (0cm) node[anchor=south]{$v_3^1$};
    \filldraw (3.2, 8.2) circle (0cm) node[anchor=south]{$u_3^1$};
    \filldraw (4.8, 8.2) circle (0cm) node[anchor=south]{$w_3^1$};
    \filldraw (5.7, 8.2) circle (0cm) node[anchor=south]{$u_3^2$};
    \filldraw (7.3, 8.2) circle (0cm) node[anchor=south]{$w_3^2$};
    \filldraw (8.2, 8.2) circle (0cm) node[anchor=south]{$u_3^3$};
    \filldraw (9.8, 8.2) circle (0cm) node[anchor=south]{$w_3^3$};
    \filldraw (7, 7.5) circle (0cm) node[anchor=south]{$v_3^2$};
    \filldraw (9.5, 7.5) circle (0cm) node[anchor=south]{$v_3^3$};
    
    \filldraw (0.75, 3.5) circle (0cm) node[anchor=south]{$v_4^0$};
    \filldraw (4.5, 3.5) circle (0cm) node[anchor=south]{$v_4^1$};
    \filldraw (3.2, 4.2) circle (0cm) node[anchor=south]{$u_4^1$};
    \filldraw (4.8, 4.2) circle (0cm) node[anchor=south]{$w_4^1$};    
    \filldraw (5.7, 4.2) circle (0cm) node[anchor=south]{$u_4^2$};
    \filldraw (7.3, 4.2) circle (0cm) node[anchor=south]{$w_4^2$};
    \filldraw (8.2, 4.2) circle (0cm) node[anchor=south]{$u_4^3$};
    \filldraw (9.8, 4.2) circle (0cm) node[anchor=south]{$w_4^3$};
    \filldraw (7, 3.5) circle (0cm) node[anchor=south]{$v_4^2$};
    \filldraw (9.5, 3.5) circle (0cm) node[anchor=south]{$v_4^3$};
    
    \filldraw (0.75, -0.5) circle (0cm) node[anchor=south]{$v_5^0$};
    \filldraw (4.5, -0.5) circle (0cm) node[anchor=south]{$v_5^1$};
    \filldraw (3.2, 0.2) circle (0cm) node[anchor=south]{$u_5^1$};
    \filldraw (4.8, 0.2) circle (0cm) node[anchor=south]{$w_5^1$};    
    \filldraw (5.7, 0.2) circle (0cm) node[anchor=south]{$u_5^2$};
    \filldraw (7.3, 0.2) circle (0cm) node[anchor=south]{$w_5^2$};
    \filldraw (8.2, 0.2) circle (0cm) node[anchor=south]{$u_5^3$};
    \filldraw (9.8, 0.2) circle (0cm) node[anchor=south]{$w_5^3$};
    \filldraw (7, -0.5) circle (0cm) node[anchor=south]{$v_5^2$};
    \filldraw (9.5, -0.5) circle (0cm) node[anchor=south]{$v_5^3$};
    
    \filldraw (0.75, -4.5) circle (0cm) node[anchor=south]{$v_6^0$};
    \filldraw (4.5, -4.5) circle (0cm) node[anchor=south]{$v_6^1$};
    \filldraw (3.2, -3.8) circle (0cm) node[anchor=south]{$u_6^1$};
    \filldraw (4.8, -3.8) circle (0cm) node[anchor=south]{$w_6^1$};    
    \filldraw (5.7, -3.8) circle (0cm) node[anchor=south]{$u_6^2$};
    \filldraw (7.3, -3.8) circle (0cm) node[anchor=south]{$w_6^2$};
    \filldraw (8.2, -3.8) circle (0cm) node[anchor=south]{$u_6^3$};
    \filldraw (9.8, -3.8) circle (0cm) node[anchor=south]{$w_6^3$};
    \filldraw (7, -4.5) circle (0cm) node[anchor=south]{$v_6^2$};
    \filldraw (9.5, -4.5) circle (0cm) node[anchor=south]{$v_6^3$};

    \filldraw (-1, 16.75) circle (0cm) node[anchor=south]{$u_1^0$};
    \filldraw (-1, 14.25) circle (0cm) node[anchor=south]{$w_1^0$};

    \filldraw (-1, 12.75) circle (0cm) node[anchor=south]{$u_2^0$};
    \filldraw (-1, 10.25) circle (0cm) node[anchor=south]{$w_2^0$};

    \filldraw (-1, 8.75) circle (0cm) node[anchor=south]{$u_3^0$};
    \filldraw (-1, 6.25) circle (0cm) node[anchor=south]{$w_3^0$};

    \filldraw (-1, 4.75) circle (0cm) node[anchor=south]{$u_4^0$};
    \filldraw (-1, 2.25) circle (0cm) node[anchor=south]{$w_4^0$};
    
    \filldraw (-1, 0.75) circle (0cm) node[anchor=south]{$u_5^0$};
    \filldraw (-1, -1.75) circle (0cm) node[anchor=south]{$w_5^0$};
    
    \filldraw (-1, -3.25) circle (0cm) node[anchor=south]{$u_6^0$};
    \filldraw (-1, -5.75) circle (0cm) node[anchor=south]{$w_6^0$};

    \filldraw (1.75, 15) circle (0cm) node[anchor=south]{$s_1$};
    \filldraw (1.75, 11) circle (0cm) node[anchor=south]{$s_2$};
    \filldraw (1.75, 7) circle (0cm) node[anchor=south]{$s_3$};
    \filldraw (1.75, 3) circle (0cm) node[anchor=south]{$s_4$};
    \filldraw (1.75, -1) circle (0cm) node[anchor=south]{$s_5$};
    \filldraw (1.75, -5) circle (0cm) node[anchor=south]{$s_6$};
    \end{tikzpicture}\begin{adjustwidth}{0pt}{0pt}
    \caption{A 6-vertex cubic graph $G$ and its corresponding $G'$}
    \label{fig:Fig2}
    \end{adjustwidth}
\end{figure}
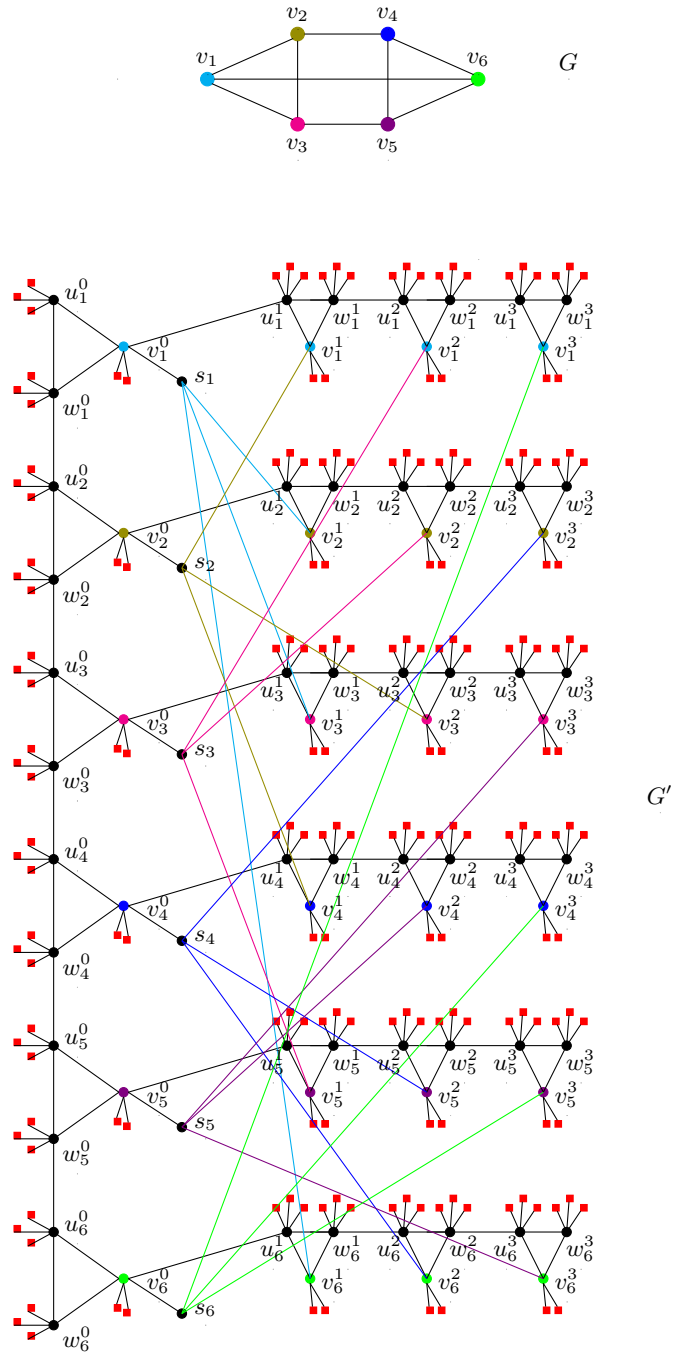
\begin{enumerate}
    \item For every vertex $v_i \in V$ in $G$, we introduce four copies in $G'$, which are represented by $v_i^0, v_i^1, v_i^2, v_i^3$.
    \item For every vertex $v_i^j \in G', 1 \leq i \leq n, 0 \leq j \leq 3$, we introduce two vertices $u_i^j, w_i^j$. We make all the three vertices $v_i^j, u_i^j$ and $w_i^j$ adjacent to each other. 
    \item We make each $v_i^0$ adjacent to $u_i^1$, each $w_i^1$ adjacent to $u_i^2$ and each $w_i^2$ adjacent to $u_i^3$. Additionally, we also make $w_i^0$ adjacent to $u_{i+1}^0$ for each $i \in \{1, ..., n-1\}$.
    \item For every vertex $v_i^0$, we introduce a vertex $s_i$, and make it adjacent to $v_i^0$. We make $s_i$ also adjacent to $v_k^j$ for all $k \in N_G(v_i)$, with the smallest possible value of $j \in \{1,2,3\}$ such that $v_k^j$ has no other $s_k$ adjacent to it. For example, $v_1$ is adjacent to $v_2, v_3$ and $v_6$ in $G$, so we make $s_1$ adjacent to $v_2^1$, $v_3^1$ and $v_6^1$ in $G'$ as all three of them do not have any other $s_k$ adjacent to them at this point. Similarly, $v_2$ is adjacent to $v_1, v_3$ and $v_4$ in $G$, we make $s_2$ adjacent to $v_1^1, v_3^2$ and $v_4^1$. Here $v_3^1$ is adjacent to $s_1$, hence we go for $v_3^2$. We add these edges lexicographically, starting from $s_1$.
    \end{enumerate}
    From hereon, we add the forbidden vertices that are constructed using the gadget.
    \begin{enumerate}
    \setcounter{enumi}{4}
    \item We make every vertex of $v_i^j, 1 \leq i \leq n, 0 \leq j \leq 3$, adjacent to two forbidden vertices.
    \item We make every vertex of $\{u_i^j \cup w_i^j\}, 1 \leq i \leq n, 0 \leq j \leq 3$, adjacent to three forbidden vertices.
\end{enumerate} \vspace{2mm}
The vertices corresponding to the missing edge in a 6-regular Ramanujan graph can be used as the forbidden vertices. We use multiple gadgets to represent all the forbidden vertices in the reduction. \vspace{3mm} \\
\noindent\textbf{Lemma 7.} If $G$ has a dominating set of size at most $k$ then $G'$ has a defensive alliance of size at most $k'$, where $k' = 4n+8k$. \vspace{3mm} \\
\textit{Proof.} Let the set $S$ be a dominating set of size at most $k$ then we claim that the set $S' = X \cup Y \cup Z$ is a defensive alliance of $G'$ of size at most $k'$, where \\
$X = \bigcup\limits_{i=1}^{n}$ $\{u_i^0 \cup v_i^0 \cup w_i^0\}$ \vspace{1mm}\\
$Y = \bigcup\limits_{i | v_i \in S} \bigcup\limits_{j \in \{1,2,3\}}$ $\{u_i^j \cup v_i^j \cup w_i^j\}$\vspace{1mm}\\
$Z = \bigcup\limits_{i | v_i \notin S} s_i$\vspace{2mm} \\
For $S'$ to be a defensive alliance, every vertex from the sets $X, Y$ and $Z$ should be protected. \vspace{2mm} \\
\textbf{Set $X$:}
\begin{itemize}
    \item Consider a vertex $v \in \{v_i^0, 1 \leq i \leq n\}$. $v$ has seven vertices in its closed neighbourhood including the two forbidden vertices. $v$ has three of its neighbours, $u_i^0, w_i^0$ and one among $u_i^1$ or $s_i$ in $S'$. Including itself, $v$ has four closed neighbours in $S'$, which makes $v$ protected. 
    \item Similarly, it can be seen that a vertex $v \in \{u_i^0 \cup w_i^0, 1 \leq i \leq n\}$ has the majority of its neighbours in $S'$. Hence, $v$ is protected.
\end{itemize}
\textbf{Set $Y$:}
\begin{itemize}
    \item Consider a vertex $v \in \{u_i^j \cup w_i^j, i | v_i\in S, 1 \leq j \leq 3\}$. $v$ is adjacent to three forbidden vertices. As all the other neighbours of $v$ are in $S'$, it is easy to verify that $v$ is protected. 
    \item A vertex $v \in \{v_i^j, i | v_i\in S, 1 \leq j \leq 3\}$ has six vertices in its closed neighbourhood which includes two forbidden vertices. As the two neighbours of $v$, $u_i^j$ and $w_i^j$ are in $S'$, $v$ is protected.
\end{itemize}
\textbf{Set $Z$:} \vspace{2mm} \\
Consider a vertex $v \in \{s_i, i | v_i \notin S\}$. $v$ has a total of five vertices in its closed neighbourhood. $v$ has two of its neighbours, $v_i^0$ and one of its other three neighbours in $S'$. Including itself, $v$ has three closed neighbours in $S'$ . Hence, $v$ is protected.  \vspace{2mm} \\
As all the vertices of the sets $X, Y$ and $Z$ are protected. As $|X| = 3n, |Y| = 9k, |Z| = n-k$ and $|X| +|Y| +|Z| = 4n +8k = k'$. Therefore, $S'$ is a defensive alliance of size at most $k'$.  This concludes the proof of Lemma 7. \qed \vspace{3mm}

\noindent\textbf{Lemma 8.}  If $G'$ has a defensive alliance of size at most $k'$ then $G$ has a dominating set of size at most $k$.\vspace{3mm} \\
\textit{Proof.} Let $S'$ be a defensive alliance in $G'$ of size at most $k'$. We define the sets $X, Y_i$ and $Z_i$ as follows. \vspace{3mm} \\
$X = \bigcup\limits_{i=1}^{n}$ $\{u_i^0 \cup v_i^0 \cup w_i^0\}$ \vspace{1mm}\\
$Y_i = \bigcup\limits_{j \in \{1,2,3\}}$ $\{u_i^j \cup v_i^j \cup w_i^j\},$ for $1 \leq i \leq n$\vspace{1mm}\\
$Z_i = s_i,$ for $1 \leq i \leq n$ \\
\begin{enumerate}
    \item If $v \in X$ is a part of the defensive alliance $S'$, then $X \subseteq S'$.
        \subitem Picking any vertex from $\{u_i^0, w_i^0\}, 1 \leq i \leq n$ would lead to $X \subseteq S'$.
        \subitem Let $v \in v_i^0, 1\leq i \leq n$ be a vertex in $S'$. In the neighbourhood of $v$, even after picking both the vertices from $Y_i, Z_i$ to be a part of $S'$, there is still a deficiency of one that needs to be filled by one among $u_i^0, w_i^0$. This creates a chain reaction that pushes all of $X$ to $S'$.
    \item If $v \in \{Y_i \cup Z_i\}$ is a part of the defensive alliance $S'$, then $X \subseteq S'$.
        \subitem Let $v \in Y_i$ is in $S'$. For $v$ to be protected, all the neighbours of $v$ (excluding the forbidden vertices) must be in $S'$. This triggers a chain reaction that pushes all of $Y_i$ and also $v_i^0$ to $S'$. As $v_i^0$ is in $S'$, $X \subseteq S'$. 
        \subitem Let $v \in Z_i$ is a part of $S'$. For $v$ to be protected, one of its neighbours from $Y_k$, $k \in N_G[v]$ must be in $S$. This would again lead to all of $Y_k$ and $v_k^0$ being in $S'$ and hence $X \subseteq S'$.
    \item It is clear that if $S'$ is non-empty then $X$ must be a part of $S'$. Here, we have consumed $3n$ vertices of $k'$ and are left with $ n + 8k$ vertices that can still go into $S'$.
    \item Note that by now, vertex $v \in \{u_i^0 \cup w_i^0\}, 1 \leq i \leq n$ is covered. Each vertex $v \in v_i^0, 1 \leq i \leq n$ needs one more neighbour either from $Y_i$ (or) $Z_i$ to be included in $S'$ for it to be protected.
        \subitem If we choose to include the vertex $u_i^1$ from $Y_i$, then this would trigger a chain reaction that pushes $Y_i$ to $S'$.
        \subitem If we choose to pick the neighbour $s_i$ from $Z_i$ then this would lead to pushing only one vertex to $S'$ before encountering a copy of $v_j$, which also needs to be a part of $S'$.
    \item We can't pick neighbours to all $v_i^0$'s from $Y_i$ as this will force us to go beyond the remaining capacity of $k'$. Based on the remaining threshold of $k'$, which is $n+8k$, it is clear that we pick neighbours from $Y_i$ to be a part of $S'$ for $k$ number of $v_i^0$'s and from $Z_i$ for $n-k$ number of $v_i^0$'s.
    \item Consider a vertex $v_i^0$ for which we pick the neighbour from $s_i$, we finally encounter a copy of $v_k$ which is an adjacent vertex of $v_i$ in $G$ and it needs to be a part of $S'$ for $v_i^0$ to be protected. This also indicates that if $v_i \notin S$ in $G$, then one of its neighbours $v_k$ must be in $S$. Basically, if $s_i \notin S'$ then the corresponding $v_i(G) \in S$ and $S$ forms a dominating set.
\end{enumerate}
Therefore, it can be inferred that the vertices part of $S$ must form a dominating set of size at most $k$. This concludes the proof of Lemma 8. \qed \vspace{3mm}
\indent In \autoref{fig:Fig2}, let $\{v_2, v_5\}$ be a dominating set of $G$ of size 2, then the corresponding defensive alliance in $G'$ is $\bigcup\limits_{i=1}^{6}$ $\{u_i^0 \cup v_i^0 \cup w_i^0\}$ $\cup \bigcup\limits_{i \in \{2, 5\}} \bigcup\limits_{j \in \{1,2,3\}}$ $\{u_i^j \cup v_i^j \cup w_i^j\}$ $\cup $ $\bigcup\limits_{i \in \{1, 3, 4, 6\}} s_i$ of size 40. \vspace{3mm} \\
\indent We have a \textit{6-regular Ramanujan graph with one missing edge} of size polynomial in $k'$. By using the vertices of the missing edge in place of the forbidden vertices, the closed neighbourhood grows to seven. Even if the new adjacent vertex from $I'$ is in the defensive alliance, we still compute MSMD(3) with one missing edge from the gadget, whose cardinality will be at least $k'+1$. \vspace{2mm} \\
\indent $G'$ has a maximum degree of six. Hence, the D\scriptsize{EFENSIVE ALLIANCE }\normalsize is NP-Complete on $\Delta(G) = 6$ graphs. This concludes the Proof of Theorem 2. \vspace{3mm} \\
\noindent\textbf{Theorem 3.} \textit{The} D\scriptsize{EFENSIVE ALLIANCE }\normalsize  \textit{problem is para-NP-hard for the parameter maximum degree ($\Delta$)}.\vspace{2mm} \\
\textit{Proof.} From Theorem 2, we have that the D\scriptsize{EFENSIVE ALLIANCE }\normalsize problem is NP-Complete on $\Delta(G) = 6$ graphs. As the problem is NP-complete for a constant value of the parameter, it implies that the D\scriptsize{EFENSIVE ALLIANCE }\normalsize problem is para-NP-hard for the parameter maximum degree ($\Delta$).
\section{D\scriptsize{EFENSIVE ALLIANCE }\normalsize parameterized by distance to clique}
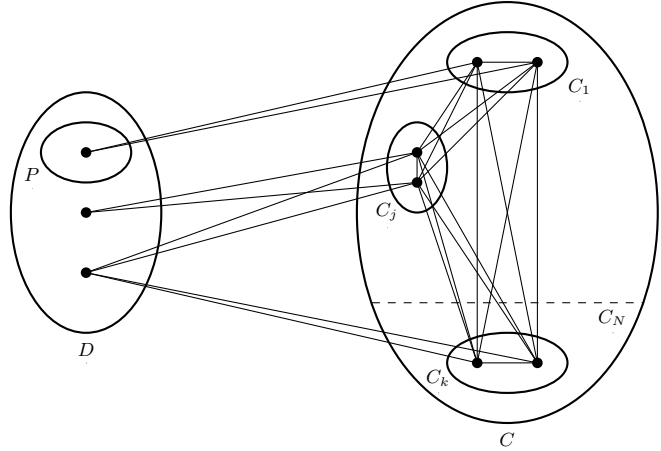
\begin{figure}
\centering
    \begin{tikzpicture} [thick,scale=0.8, every node/.style={scale=0.8}]
        \draw (0, 0) ellipse (1.25 and 2);
        \filldraw (0, -2.5) circle (0cm) node[anchor=south]{$D$};
        \draw (0, 1) ellipse (0.75 and 0.5);
        \filldraw (-0.9, 0.4) circle (0cm) node[anchor=south]{$P$};
        \draw (7,0) ellipse (2.5 and 3.5);
        \filldraw (7, -4) circle (0cm) node[anchor=south]{$C$};
        \draw (7, 2.5) ellipse (1 and 0.5);
        \filldraw (8.2, 1.85) circle (0cm) node[anchor=south]{$C_1$};
        \draw (5.5, 0.75) ellipse (0.5 and 0.75);
        \filldraw (5, -0.25) circle (0cm) node[anchor=south]{$C_j$};
        \draw (7, -2.5) ellipse (1 and 0.5);
        \filldraw (5.85, -3) circle (0cm) node[anchor=south]{$C_k$};
        \filldraw (8.75, -2) circle (0cm) node[anchor=south]{$C_N$};
        \filldraw (0, 1) circle (2pt) node[anchor=south]{};
        \filldraw (0, 0) circle (2pt) node[anchor=south]{};
        \filldraw (0, -1) circle (2pt) node[anchor=south]{};        
        \filldraw (6.5, 2.5) circle (2pt) node[anchor=south]{};
        \filldraw (7.5, 2.5) circle (2pt) node[anchor=south]{};   
        \filldraw (5.5, 0.5) circle (2pt) node[anchor=south]{};
        \filldraw (5.5, 1) circle (2pt) node[anchor=south]{};
        \filldraw (6.5, -2.5) circle (2pt) node[anchor=south]{};
        \filldraw (7.5, -2.5) circle (2pt) node[anchor=south]{};
        \draw[thin] (0, 1) -- (6.5, 2.5);
        \draw[thin] (0, 1) -- (7.5, 2.5);
        \draw[thin] (0, 0) -- (5.5, 0.5);
        \draw[thin] (0, 0) -- (5.5, 1);        
        \draw[thin] (0, -1) -- (5.5, 0.5);
        \draw[thin] (0, -1) -- (5.5, 1);
        \draw[thin] (0, -1) -- (6.5, -2.5);
        \draw[thin] (0, -1) -- (7.5, -2.5);
        \draw[thin] (6.5, 2.5) -- (7.5, 2.5);
        \draw[thin] (6.5, 2.5) -- (5.5, 0.5);
        \draw[thin] (6.5, 2.5) -- (5.5, 1);
        \draw[thin] (6.5, 2.5) -- (6.5, -2.5);
        \draw[thin] (6.5, 2.5) -- (7.5, -2.5);
        \draw[thin] (7.5, 2.5) -- (5.5, 0.5);
        \draw[thin] (7.5, 2.5) -- (5.5, 1);
        \draw[thin] (7.5, 2.5) -- (6.5, -2.5);
        \draw[thin] (7.5, 2.5) -- (7.5, -2.5);
        \draw[thin] (5.5, 0.5) -- (5.5, 1);
        \draw[thin] (5.5, 0.5) -- (6.5, -2.5);
        \draw[thin] (5.5, 0.5) -- (7.5, -2.5);
        \draw[thin] (5.5, 1) -- (6.5, -2.5);
        \draw[thin] (5.5, 1) -- (7.5,-2.5);
        \draw[thin] (6.5, -2.5) -- (7.5, -2.5);
        \draw[thin, dashed] (4.75, -1.5) -- (9.25, -1.5);
    \end{tikzpicture}
    \caption{Partitioning of the vertex set $V$ into sets $D$ and $C$, where $|D|$ is the distance to clique and $C$ is a clique.} 
    \label{fig:Fig3}
\end{figure}

In this section, we show that the D\scriptsize{EFENSIVE ALLIANCE }\normalsize problem is FPT parameterized by distance to clique. We reduce the given problem to the integer linear programming problem (ILP), which is known to be FPT when parameterized by the number of variables. \vspace{3mm}  \\
\textbf{Definition 2.} For a graph $G = (V, E)$, the parameter \textit{distance to clique} is the cardinality of the smallest set $D \subseteq V$ such that $V\setminus D$ is a clique.\vspace{2mm} \\
We can use a simple branching algorithm to compute set $D$ of size at most $k$ in $\mathcal{O}^*(2^k)$ time, if such a set exists. \vspace{3mm} \\
\noindent\textbf{\normalsize ILP formulation}\vspace{2mm}\\
Integer Linear Programming is a framework used to formulate a given problem using a finite number of variables. The problem definition is given as follows: \vspace{3mm} \\
\textbf{Problem. }$p$-Opt-ILP 
\begin{adjustwidth}{1em}{0pt}
\textit{Instance:} A matrix $A \in \mathbb{Z}^{m*p}$, and vectors $b \in \mathbb{Z}^m$ and $c \in \mathbb{Z}^p$.\\
\textit{Objective: }Find a vector $x \in \mathbb{Z}^p$ that minimizes $c^\top x$ and satisfies that $Ax \geq b$.\\
\textit{Parameter: }$p$, the number of variables. \\
\end{adjustwidth}
Lenstra~\cite{LNT} showed that deciding the feasibility of a $p$-ILP is fixed-parameter tractable with running time doubly exponential in $p$, where $p$ is the number of variables. Later, Kannan~\cite{KNN} gave a $p^p$ algorithm for $p$-ILP.  Fellows et al.~\cite{FLWS} proved that $p$-Opt-ILP, the optimization version of the problem, is also fixed-parameter tractable. \vspace{3mm} \\
\textbf{Theorem 4.}~\cite{FLWS} The optimization version of $p$-variable I\scriptsize{NTEGER} \normalsize L\scriptsize{INEAR} \normalsize P\scriptsize{ROGRAMMING }\normalsize can be solved using $\mathcal{O}(p^{2.5p+o(p)} \cdot L \cdot log(MN))$ arithmetic operations and space
polynomial in $L$, where $L$ is the number of bits in the input, $N$ is the maximum absolute value any variable can take, and $M$ is an
upper bound on the absolute value of the minimum taken by the objective function.\vspace{3mm} \\
\noindent \textbf{Theorem 5.} Given a graph $G = (V, E)$ and $D \subseteq V$ such that $V\setminus D$ is a clique, the D\scriptsize{EFENSIVE ALLIANCE }\normalsize problem can be solved in $\mathcal{O}^*(f(|D|))$ time.\vspace{3mm} \\
Consider a graph $G = (V, E)$ and $D \subseteq V$ such that $|D|$ is the distance to clique of $G$ and $C = V \setminus D$. We partition the vertices of $C$ into $t$ twin classes, which are represented by $C_1,C_2 ,...,C_t$ ($t \leq 2^{|D|})$, such that all the vertices in a twin class $C_i$ have same adjacency in $D$. Let $S \subseteq V$ be a defensive alliance of $G$. We guess the vertex sets $P = S \cap D$ and compute $S_C = S \cap C$. We also guess a subset of twin classes $C_{N} \subseteq C$ from which no vertices are picked in the solution. See \autoref{fig:Fig3} for an illustration. After the guess of $P$ and $C_N$, we compute $S_C$ using integer linear programming. For each of $u \in D$, we define $demand(u)$ = $\lceil\frac{1}{2}(d(u)+1)\rceil-|N[u] \cap P|$. For each $u \in D$, we denote by $M(u)$ the set of indices $i$ such that $C_i \subseteq N(u)$. In other words, $M(u)$ represents the indices of the twin classes from $C$ that $u$ is adjacent to. Let $x_i$ represent the number of vertices in $C_i \cap S$. In our ILP formulation, there are $t$ variables that are $x_1, x_2, ..., x_t$. \vspace{3mm} \\
\setlength{\textfloatsep}{1\baselineskip plus 0\baselineskip minus 0\baselineskip}
\textbf{Lemma 9.} The set $S$ is a defensive alliance if and only if
\begin{enumerate}
\item For $\forall C_i\in C_N$, $x_i$ = 0.
\item For each $u \in P$, $ \sum_{i\in M(u)}^{}x_i \geq$ $demand(u)$.
\item Each $v \in C \setminus C_N$ has to satisfy $|N(v) \cap P| + \sum_{i \in \{1,2,...,t\}}^{} x_i \geq |N[v] \setminus P|  - \sum_{i \in \{1,2,...,t\}}^{} x_i -1.$
\end{enumerate}
\textit{Proof.}
\begin{enumerate}
\item As $C_N$ represents the set of twin classes with no vertices in $S$, we have $x_i$ = 0 for all $C_i \in C_N$.
\item For each $u \in P$, $d(u)$ = $|N(u) \cap S| + |N(u) \setminus S|$ and $|N(u) \cap S| \geq |N(u)\setminus S|-1$ holds if and only if $2 * |N(u) \cap S| \geq d(u)-1$, which is equivalent to $|N(u) \cap S_C| \geq demand(u)$ implies $\sum_{i\in M(u)}^{}x_i \geq$ $demand(u)$.
\item For each $v \in C \setminus C_N$, $|N(v) \cap S|$ = $|N(v) \cap P| + |S_C|$, which is indeed $|N(v) \cap P| + \sum_{i \in \{1,2,...,t\}}^{} x_i$; $|N(v)\setminus S|-1$ = $|(N(v) \cap D) \setminus P|+ |C| - \sum_{i \in \{1,2,...,t\}}^{} x_i -1$, which equals $|N[v] \setminus P|  - \sum_{i \in \{1,2,...,t\}}^{} x_i -1.$ \qed
\end{enumerate}
\noindent The ILP formulation for the defensive alliance is given as \vspace{2mm}\\
\noindent\fbox{%
    \parbox{\textwidth}{%
    Minimize $\sum_{i \in \{1,2,...,t\}}^{}x_i$ \vspace{2mm} \\
    Subject to
    \begin{itemize}
    \item $x_i = 0$, for $\forall C_i \in C_N$.
    \item $\sum_{i\in M(u)}^{}x_i \geq$ $demand(u)$, for each $u \in P.$
    \item $|N(v) \cap P| + \sum_{i \in \{1,2,...,t\}}^{} x_i \geq |N[v] \setminus P|  - \sum_{i \in \{1,2,...,t\}}^{} x_i -1,$ for every $v \in C \setminus C_N.$
    \item $x_i \leq |C_i|$, for each $i \in \{1,2,..., t\}.$
    \end{itemize}
    }%
}
\vspace{3mm} \\
The ILP will output the optimal values of $x_i$ for all $i \in \{1,2,...,t\}$. If $x_i > 0$, we need to pick $x_i$ vertices from $C_i$. As all the vertices in $C_i$ have the same neighbourhood, we can pick any $x_i$ vertices. Hence, we obtain the vertex set $S_C$. \vspace{3mm} \\
In our ILP formulation, we have at most $2^{|D|}$ variables. The values of all the variables and the objective function are bounded by $n$. The constraints can be represented using $\mathcal{O}(4^{|D|} \cdot log n)$ bits. With the help of Theorem 4, we will be able to solve the problem with the guess ($P,\; C_N$) in FPT time. There are $2^{|D|}$ candidates for $P$ and $2^{2^{|D|}}$ candidates for $C_N$. To obtain $S_C$, we solve $8^{|D|}$ ILP formulas where each formula can be computed in $\mathcal{O}^*(f(|D|))$ time. This concludes the proof of Theorem 5.

\section{D\footnotesize{EFENSIVE ALLIANCE }\normalsize parameterized by twin cover}
In this section, we show that the D\scriptsize{EFENSIVE ALLIANCE }\normalsize problem is FPT parameterized by twin cover. We give an ILP formulation for the combined parameter twin cover, the size of the largest clique outside the twin cover. Using this, we show that the D\scriptsize{EFENSIVE ALLIANCE }\normalsize problem is FPT for the parameter twin cover. \vspace{3mm}  \\
\textbf{Definition 3.} For a graph $G = (V, E)$, the parameter \textit{twin cover} is the cardinality of the smallest set $T \subseteq V$ such that $V\setminus T$ is a disjoint union of cliques wherein all the vertices in each clique have the same adjacency in the twin cover.\vspace{2mm} \\
\textbf{Theorem 6.}~\cite{GNN} If a minimum twin cover in $G$ has size at most $k$, then it is possible to compute a twin cover of size at most $k$ in time $\mathcal{O}(|E||V|+k|V|+1.2738^k)$. \vspace{2mm} \\
\textbf{Theorem 7.} Given a graph $G = (V, E)$, $T \subseteq V$ is a twin cover of $G$ and $z$ is the size of the largest clique outside the twin cover, the D\scriptsize{EFENSIVE ALLIANCE }\normalsize problem can be solved in $\mathcal{O}^*(f(|T|, z))$ time.\vspace{3mm} \\\setlength{\textfloatsep}{1\baselineskip plus 0\baselineskip minus 0\baselineskip}
\begin{figure}
    \centering
    \begin{tikzpicture} [thick,scale=0.7, every node/.style={scale=0.7}]
        \draw (-9, -8) ellipse (1.5 and 4);
        \draw (-9, -6) ellipse (0.6 and 1);
        \filldraw [blue] (-9, -6) circle (2.5pt);
        \filldraw [black] (-9, -8.25) circle (2.5pt);
        \filldraw [black] (-9, -10.5) circle (2.5pt);
        \draw[thin, dashed] (-9, -6) -- (-7, 0);
        \draw[thin, dashed] (-7, 0) -- (-3.4, 1.85);
        \draw[thin, dashed] (-7, 0) -- (-3.4, -1.85);
        \draw[thin, dashed] (-9, -6) -- (-7, -6);
        \draw[thin, dashed] (-9, -8.25) -- (-7, -6);
        \draw[thin, dashed] (-7, -6) -- (-3.4, -4.15);
        \draw[thin, dashed] (-7, -6) -- (-3.4, -7.85);
        \draw[thin, dashed] (-9, -10.5) -- (-7, -16);
        \draw[thin, dashed] (-7, -16) -- (-3.4, -14.15);
        \draw[thin, dashed] (-7, -16) -- (-3.4, -17.85);
        \node[] at (-9.9, -6.5) {$P$};
        \node[] at (-9, -12.5) {$T$};
        
        \draw (0, 0) ellipse (5 and 2.5);
        \filldraw (5.5, 0) circle (0cm) node[anchor=south]{$C_1$};
        \draw[thin] (-4,1.5) -- (-4,-1.5);
        \draw[thin] (0,2.5) -- (0,-2.5);
        \draw[thin] (3,2) -- (3,-2);
        \draw[thin] (-4, 0.8) -- (0, 0.8);
        \draw[thin] (-4, -0.8) -- (0, -0.8);
        \draw[thin] (-3.5, 1) -- (-3.5, 1.6);
        \draw[thin] (-3, 1) -- (-3, 1.6);
        \draw[thin] (-2.5, 1) -- (-2.5, 1.6);
        \draw[thin] (-0.5, 1) -- (-0.5, 1.6);
        \draw[thin] (-3.5, -0.2) -- (-3.5, 0.4);
        \draw[black, thin] (-3, -0.2) -- (-3, 0.4);
        \draw[black, thin] (-2.5, -0.2) -- (-2.5, 0.4);
        \draw[black, thin] (-0.5, -0.2) -- (-0.5, 0.4);
        \draw[black, thin] (-3.5, -1) -- (-3.5, -1.6);
        \draw[black, thin] (-3, -1) -- (-3, -1.6);
        \draw[black, thin] (-2.5, -1) -- (-2.5, -1.6);
        \draw[black, thin] (-0.5, -1) -- (-0.5, -1.6);
        \filldraw [blue] (-3.5, 1) circle (2.5pt);
        \filldraw [blue] (-3, 1) circle (2.5pt);
        \filldraw [blue] (-2.5, 1) circle (2.5pt);
        \filldraw [blue] (-0.5, 1) circle (2.5pt);
        \filldraw [blue] (-3.5, 1.6) circle (2.5pt);
        \filldraw [blue] (-3, 1.6) circle (2.5pt);
        \filldraw [blue] (-2.5, 1.6) circle (2.5pt);
        \filldraw [blue] (-0.5, 1.6) circle (2.5pt);
        \filldraw [black] (-3.5, -0.2) circle (2.5pt);
        \filldraw [blue] (-3, -0.2) circle (2.5pt);
        \filldraw [black] (-2.5, -0.2) circle (2.5pt);
        \filldraw [blue] (-0.5, -0.2) circle (2.5pt);
        \filldraw [black] (-3.5, -1) circle (2.5pt);
        \filldraw [black] (-3, -1) circle (2.5pt);
        \filldraw [black] (-2.5, -1) circle (2.5pt);
        \filldraw [black] (-0.5, -1) circle (2.5pt);
        
        \filldraw [gray] (-1.75, 1.3) circle (1.5pt);
        \filldraw [gray] (-1.5, 1.3) circle (1.5pt);
        \filldraw [gray] (-1.25, 1.3) circle (1.5pt);
        \filldraw [gray] (-1.75, 0.1) circle (1.5pt);
        \filldraw [gray] (-1.5, 0.1) circle (1.5pt);
        \filldraw [gray] (-1.25, 0.1) circle (1.5pt);
        \filldraw [gray] (-1.75, -1.3) circle (1.5pt);
        \filldraw [gray] (-1.5, -1.3) circle (1.5pt);
        \filldraw [gray] (-1.25, -1.3) circle (1.5pt);
        
        \filldraw [black] (-3, 0.4) circle (2.5pt);
        \filldraw [blue] (-2.5, 0.4) circle (2.5pt);
        \filldraw [black] (-0.5, 0.4) circle (2.5pt);
        \filldraw [blue] (-3.5, 0.4) circle (2.5pt);
        \filldraw [black] (-3.5, -1.6) circle (2.5pt);
        \filldraw [black] (-3, -1.6) circle (2.5pt);
        \filldraw [black] (-2.5, -1.6) circle (2.5pt);
        \filldraw [black] (-0.5, -1.6) circle (2.5pt);
        
        \filldraw [gray] (1.3, 0) circle (1.5pt);
        \filldraw [gray] (1.6, 0) circle (1.5pt);
        \filldraw [gray] (1.9, 0) circle (1.5pt);
        \node[] at (-4.5,0) {$C_1^1$};
        \node[] at (-0.5,2.1) {$C_1^{2, F}$};
        \node[] at (-0.5,-2.1) {$C_1^{2, N}$};
        \node[] at (-3.6,-0.5) {\tiny \scalebox{0.8}{$C_1^{2, P_1}$}};
        \node[] at (-2.9,-0.5) {\tiny \scalebox{0.8}{$C_1^{2, P_2}$}};
        \node[] at (-2.2,-0.5) {\tiny \scalebox{0.8}{$C_1^{2, P_3}$}};
        \node[] at (-0.5,-0.52) {\tiny \scalebox{0.75}{$C_1^{2, P_{y_1^2}}$}};
        \node[] at (4,0) {$C_1^z$};

        \draw (0, -6) ellipse (5 and 2.5);
        \filldraw (5.5, -6) circle (0cm) node[anchor=south]{$C_2$};
        \draw[thin] (-4,-4.5) -- (-4,-7.5);
        \draw[thin] (0,-3.5) -- (0,-8.5);
        \draw[thin] (3,-4) -- (3,-8);
        \draw[thin] (-4, -5.2) -- (0, -5.2);
        \draw[thin] (-4, -6.8) -- (0, -6.8);
        \draw[thin] (-3.5, -5) -- (-3.5, -4.4);
        \draw[thin] (-3, -5) -- (-3, -4.4);
        \draw[thin] (-2.5, -5) -- (-2.5, -4.4);
        \draw[thin] (-0.5, -5) -- (-0.5, -4.4);
        \draw[thin] (-3.5, -6.2) -- (-3.5, -5.6);
        \draw[black, thin] (-3, -6.2) -- (-3, -5.6);
        \draw[black, thin] (-2.5, -6.2) -- (-2.5, -5.6);
        \draw[black, thin] (-0.5, -6.2) -- (-0.5, -5.6);
        \draw[black, thin] (-3.5, -7) -- (-3.5, -7.6);
        \draw[black, thin] (-3, -7) -- (-3, -7.6);
        \draw[black, thin] (-2.5, -7) -- (-2.5, -7.6);
        \draw[black, thin] (-0.5, -7) -- (-0.5, -7.6);
        \filldraw [blue] (-3.5, -5) circle (2.5pt);
        \filldraw [blue] (-3, -5) circle (2.5pt);
        \filldraw [blue] (-2.5, -5) circle (2.5pt);
        \filldraw [blue] (-0.5, -5) circle (2.5pt);
        \filldraw [blue] (-3.5, -4.4) circle (2.5pt);
        \filldraw [blue] (-3, -4.4) circle (2.5pt);
        \filldraw [blue] (-2.5, -4.4) circle (2.5pt);
        \filldraw [blue] (-0.5, -4.4) circle (2.5pt);
        \filldraw [blue] (-3.5, -6.2) circle (2.5pt);
        \filldraw [black] (-3, -6.2) circle (2.5pt);
        \filldraw [black] (-2.5, -6.2) circle (2.5pt);
        \filldraw [blue] (-0.5, -6.2) circle (2.5pt);
        \filldraw [black] (-3.5, -7) circle (2.5pt);
        \filldraw [black] (-3, -7) circle (2.5pt);
        \filldraw [black] (-2.5, -7) circle (2.5pt);
        \filldraw [black] (-0.5, -7) circle (2.5pt);
        
        \filldraw [gray] (-1.75, -4.7) circle (1.5pt);
        \filldraw [gray] (-1.5, -4.7) circle (1.5pt);
        \filldraw [gray] (-1.25, -4.7) circle (1.5pt);
        \filldraw [gray] (-1.75, -5.9) circle (1.5pt);
        \filldraw [gray] (-1.5, -5.9) circle (1.5pt);
        \filldraw [gray] (-1.25, -5.9) circle (1.5pt);
        \filldraw [gray] (-1.75, -7.3) circle (1.5pt);
        \filldraw [gray] (-1.5, -7.3) circle (1.5pt);
        \filldraw [gray] (-1.25, -7.3) circle (1.5pt);
        
        \filldraw [blue] (-3, -5.6) circle (2.5pt);
        \filldraw [blue] (-2.5, -5.6) circle (2.5pt);
        \filldraw [black] (-0.5, -5.6) circle (2.5pt);
        \filldraw [black] (-3.5, -5.6) circle (2.5pt);
        \filldraw [black] (-3.5, -7.6) circle (2.5pt);
        \filldraw [black] (-3, -7.6) circle (2.5pt);
        \filldraw [black] (-2.5, -7.6) circle (2.5pt);
        \filldraw [black] (-0.5, -7.6) circle (2.5pt);
        
        \filldraw [gray] (1.3, -6) circle (1.5pt);
        \filldraw [gray] (1.6, -6) circle (1.5pt);
        \filldraw [gray] (1.9, -6) circle (1.5pt);
        \node[] at (-4.5,-6) {$C_2^1$};
        \node[] at (-0.5,-3.9) {$C_2^{2, F}$};
        \node[] at (-0.5,-8.1) {$C_2^{2, N}$};
        \node[] at (-3.6,-6.5) {\tiny \scalebox{0.8}{$C_2^{2, P_1}$}};
        \node[] at (-2.9,-6.5) {\tiny \scalebox{0.8}{$C_2^{2, P_2}$}};
        \node[] at (-2.2,-6.5) {\tiny \scalebox{0.8}{$C_2^{2, P_3}$}};
        \node[] at (-0.5,-6.52) {\tiny \scalebox{0.75}{$C_2^{2, P_{y_2^2}}$}};
        \node[] at (4,-6) {$C_2^z$};

        \draw (0, -16) ellipse (5 and 2.5);
        \filldraw (5.5, -16) circle (0cm) node[anchor=south]{$C_t$};
        \draw[thin] (-4,-14.5) -- (-4,-17.5);
        \draw[thin] (0,-13.5) -- (0,-18.5);
        \draw[thin] (3,-14) -- (3,-18);
        \draw[thin] (-4, -15.2) -- (0, -15.2);
        \draw[thin] (-4, -16.8) -- (0, -16.8);
        \draw[thin] (-3.5, -15) -- (-3.5, -14.4);
        \draw[thin] (-3, -15) -- (-3, -14.4);
        \draw[thin] (-2.5, -15) -- (-2.5, -14.4);
        \draw[thin] (-0.5, -15) -- (-0.5, -14.4);
        \draw[thin] (-3.5, -16.2) -- (-3.5, -15.6);
        \draw[black, thin] (-3, -16.2) -- (-3, -15.6);
        \draw[black, thin] (-2.5, -16.2) -- (-2.5, -15.6);
        \draw[black, thin] (-0.5, -16.2) -- (-0.5, -15.6);
        \draw[black, thin] (-3.5, -17) -- (-3.5, -17.6);
        \draw[black, thin] (-3, -17) -- (-3, -17.6);
        \draw[black, thin] (-2.5, -17) -- (-2.5, -17.6);
        \draw[black, thin] (-0.5, -17) -- (-0.5, -17.6);
        \filldraw [blue] (-3.5, -15) circle (2.5pt);
        \filldraw [blue] (-3, -15) circle (2.5pt);
        \filldraw [blue] (-2.5, -15) circle (2.5pt);
        \filldraw [blue] (-0.5, -15) circle (2.5pt);
        \filldraw [blue] (-3.5, -14.4) circle (2.5pt);
        \filldraw [blue] (-3, -14.4) circle (2.5pt);
        \filldraw [blue] (-2.5, -14.4) circle (2.5pt);
        \filldraw [blue] (-0.5, -14.4) circle (2.5pt);
        \filldraw [blue] (-3.5, -16.2) circle (2.5pt);
        \filldraw [black] (-3, -16.2) circle (2.5pt);
        \filldraw [blue] (-2.5, -16.2) circle (2.5pt);
        \filldraw [black] (-0.5, -16.2) circle (2.5pt);
        \filldraw [black] (-3.5, -17) circle (2.5pt);
        \filldraw [black] (-3, -17) circle (2.5pt);
        \filldraw [black] (-2.5, -17) circle (2.5pt);
        \filldraw [black] (-0.5, -17) circle (2.5pt);
        
        \filldraw [gray] (-1.75, -14.7) circle (1.5pt);
        \filldraw [gray] (-1.5, -14.7) circle (1.5pt);
        \filldraw [gray] (-1.25, -14.7) circle (1.5pt);
        \filldraw [gray] (-1.75, -15.9) circle (1.5pt);
        \filldraw [gray] (-1.5, -15.9) circle (1.5pt);
        \filldraw [gray] (-1.25, -15.9) circle (1.5pt);
        \filldraw [gray] (-1.75, -17.3) circle (1.5pt);
        \filldraw [gray] (-1.5, -17.3) circle (1.5pt);
        \filldraw [gray] (-1.25, -17.3) circle (1.5pt);
        
        \filldraw [blue] (-3, -15.6) circle (2.5pt);
        \filldraw [black] (-2.5, -15.6) circle (2.5pt);
        \filldraw [blue] (-0.5, -15.6) circle (2.5pt);
        \filldraw [black] (-3.5, -15.6) circle (2.5pt);
        \filldraw [black] (-3.5, -17.6) circle (2.5pt);
        \filldraw [black] (-3, -17.6) circle (2.5pt);
        \filldraw [black] (-2.5, -17.6) circle (2.5pt);
        \filldraw [black] (-0.5, -17.6) circle (2.5pt);
        
        \filldraw [gray] (1.3, -16) circle (1.5pt);
        \filldraw [gray] (1.6, -16) circle (1.5pt);
        \filldraw [gray] (1.9, -16) circle (1.5pt);
        \node[] at (-4.5,-16) {$C_t^1$};
        \node[] at (-0.5,-13.9) {$C_t^{2, F}$};
        \node[] at (-0.5,-18.1) {$C_t^{2, N}$};
        \node[] at (-3.6,-16.5) {\tiny \scalebox{0.8}{$C_t^{2, P_1}$}};
        \node[] at (-2.9,-16.5) {\tiny \scalebox{0.8}{$C_t^{2, P_2}$}};
        \node[] at (-2.2,-16.5) {\tiny \scalebox{0.8}{$C_t^{2, P_3}$}};
        \node[] at (-0.5,-16.52) {\tiny \scalebox{0.75}{$C_t^{2, P_{y_t^2}}$}};
        \node[] at (4,-16) {$C_t^z$};
        \node[] at (0, -19.25) {$C$};

        \filldraw [gray] (0, -10.7) circle (1.5pt);
        \filldraw [gray] (0, -11) circle (1.5pt);
        \filldraw [gray] (0, -11.3) circle (1.5pt);
    \end{tikzpicture}
    \caption{Partitioning of the vertex set $V$ into sets $T$ and $C$, where $T$ is the twin cover and $C$ is the union of clique sets outside $T$. The figure also highlights the cliques of type \textit{full}, \textit{partial} and \textit{null} of length two in each clique set. The blue vertices belong in the alliance set.} 
    \label{fig:Fig4}
\end{figure}
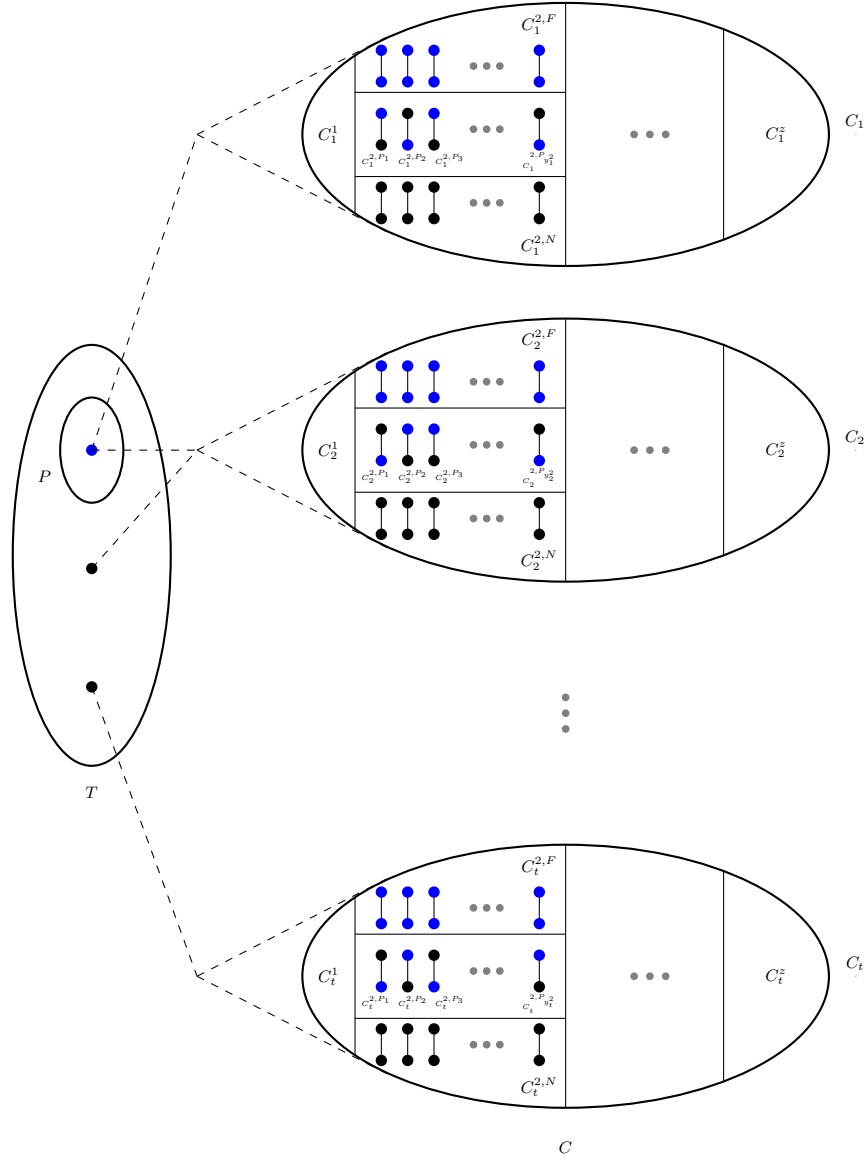
Consider a graph $G = (V, E)$. Let $T \subseteq V$ be a twin cover of $G$ and $C = V \setminus T$. We partition the vertices of $C$ into $t$ clique sets which are represented by $C_1,C_2 ,...,C_t$ ($t \leq 2^{|T|})$, such that all the vertices in a clique set $C_i$ have same adjacency in $T$. Let $S \subseteq V$ be a defensive alliance of $G$. We guess the vertex set $P = S \cap T$ and compute $S_C = S \cap C$. For each of $u \in P$, we define $demand(u)$ = $\lceil\frac{1}{2}(d(u)+1)\rceil-|N[u] \cap P|$. For each $u \in P$, we denote by $M(u)$ the set of indices $i$ such that $C_i \subseteq N(u)$. In other words, $M(u)$ represents the indices of the clique sets from $C$ that $u$ is adjacent to. We have at most $z$ different size cliques in each clique set whose sizes range from 1 to $z$. We represent the cliques of size $l$ in the clique set $C_i$ as $C_i^l$. \vspace{2mm} \\ 
\indent We place cliques of all sizes from each clique set into one of the following three types: \textit{full}, \textit{partial} and \textit{null}. \textit{full} clique has all of its vertices picked in the solution, \textit{partial} clique has some of its vertices picked, whereas a \textit{null} clique has no vertices picked. $C_i^{l, F}$ represents the union of all \textit{full} cliques in $C_i^l$. $C_i^{l, P}, C_i^{l, N}$ represent the union of all \textit{partial} cliques and union of all \textit{null} cliques in $C_i^l$ respectively. We denote each \textit{partial} clique in $C_i^{l, P}$ by $C_i^{l, P_j}$, where $j$ denotes the index of a partial clique. See \autoref{fig:Fig4} for an illustration of \textit{full}, \textit{partial} and \textit{null} cliques of length two. 
Let $x_i^{l, P_j}$ represent the number of vertices in $C_i^{l, P_j} \cap S$. In the ILP formulation, we need an individual variable for every \textit{partial} clique in $C_i^{l, P}$. Here, the idea is to limit the number of partial cliques in each clique set $C_i$, which results in formulating the ILP using the desired number of variables. \vspace{3mm} \\
\noindent \textbf{Lemma 10.} Consider a set $C_i^l$ from a clique set $C_i$, there exists an optimal solution with at most $l-1$ \textit{partial} cliques from $C_i^l$. \vspace{2mm} \\
\textit{Proof. }Consider an optimal solution $S$ with $p$ partial cliques from $C_i^l$. We construct another optimal solution $S'$ from $S$ with at most $l-1$ \textit{partial} cliques as follows. We arrange the \textit{partial} cliques in $S$ in ascending order based on their number of vertices in $S$. Let the order be $C_i^{l, P_1}, C_i^{l, P_2}, ..., C_i^{l, P_p}$. We perform the following operation repeatedly to obtain $S'$. In any iteration, let the first clique in the list be $C^*$. We replace each vertex of $C^*$ that is in $S$ with a vertex that is not in $S$ in each of the last $|C^*|$ cliques in the list. We push $C^*$ to \textit{null} clique set. If there are any other cliques that become \textit{full} by the recent addition of vertices, then we simply move them into \textit{full} cliques set. We perform this until we can place all the vertices of $C^*$ into the cliques of larger sizes. It is easy to see that all the vertices in the resultant instance are protected as the number of vertices from any clique that belongs to the solution only grows. There is no change in the number of vertices that go into the solution, therefore $S'$ is also optimal. The maximum number of \textit{partial} cliques that would remain when this process comes to a halt is $l$-1 \textit{partial} cliques with $l-1$ vertices in $S$ from each clique. Hence, we conclude that there exists an optimal solution with at most $l-1$ \textit{partial} cliques.\qed \vspace{3mm}
\noindent Let $y_i^l$ be the number of \textit{partial} cliques in $C_i^l$. From Lemma 10, it is clear that there is an optimal solution with at most $l-1$ \textit{partial} cliques from $C_i^l$ and we have $y_i^l \leq l-1$. We guess the cliques from $C_i^l$ that go into $C_i^{l, P}$ in $y_i^l+1$ ways and from the remaining $|C_i-C_i^{l, P}|$ cliques, $C_i^{l, F}$ can be guessed in $m-y_i^l+1$ ways, where $m$ is the number of cliques in $C_i^l$. As we have guessed $P, C_i^{l, F}$ and $C_i^{l, P}$, we compute $S_C$ using integer linear programming.
$x_i^{l, P}$ represents the sum of $x_i^{l, P_1}, x_i^{l, P_2},..., x_i^{l, P_{y_i^l}}$.
\noindent In our ILP formulation,
there are at most $\sum_{i = 1}^{t} \sum_{l = 1}^{z} y_i^l$ variables that are $x_1^{1, P_1}, ..., x_1^{z, P_{y_1^z}}, x_2^{1, P_1}, ...,x_2^{z, P_{y_2^z}}, ..., x_{t}^{z, P_{y_t^z}}$. \vspace{3mm} \\
\textbf{Lemma 11.} The set $S$ is a D\scriptsize{EFENSIVE ALLIANCE }\normalsize if and only if
\begin{enumerate}
\item For each $u \in P$, $\sum_{i \in M(u)}^{} $ $\sum_{l=1}^{l=z}|C_i^{l, F}|$ + $\sum_{i \in M(u)}^{}$ $\sum_{l=1}^{l=z}$ $\sum_{j=1}^{j=y_i^l} x_i^{l, P_j}$ $\geq$ $demand(u)$.
\item Each $v \in C_i^{l, P_j}$ has to satisfy $|N(v) \cap P| + x_i^{l, P_j} \geq |N[v] \setminus  P| -x_i^{l, P_j}-1.$
\end{enumerate}
\textit{Proof. }
\begin{enumerate}
\item For each $u \in P$, $d(u)$ = $|N(u) \cap S| + |N(u) \setminus S|$ and $N(u \cap S_C) = $ $\sum_{i \in M(u)}^{}$ $\sum_{l=1}^{l=z}|C_i^{l, F}| +$ $ \sum_{i \in M(u)}^{} \sum_{l=1}^{l=z} $ $\sum_{j=1}^{j=y_i^l} x_i^{l, P_j}$. $|N(u) \cap S| \geq |N(u)\setminus S|-1$ holds if and only if $2 * |N(u) \cap S| \geq$ $d(u)-1$, which implies $\sum_{i \in M(u)}^{} \sum_{l=1}^{l=z}|C_i^{l, F}|$ + $\sum_{i \in M(u)}^{} \sum_{l=1}^{l=z} \sum_{j=1}^{j=y_i^l} x_i^{l, P_j}$ $\geq$ $demand(u)$.
\item For each $v \in C_i^{l, P_j}$, $|N(v) \cap S|$ = $|N(v) \cap P| + x_i^{l, P_j}$; $|N(v)\setminus S|-1$ = $|N[v] \setminus  P| -x_i^{l, P_j}-1.$ \qed
\end{enumerate}
\noindent The ILP formulation for the D\scriptsize{EFENSIVE ALLIANCE }\normalsize is given as \vspace{2mm} \\ %
\noindent\fbox{%
    \parbox{\textwidth}{%
    Minimize $\sum_{i = 1}^{i=t} \sum_{l=1}^{l=z} \sum_{j=1}^{j=y_i^l} x_i^{l, P_j}$ \vspace{2mm} \\
    Subject to
    \begin{itemize}
    \item $\sum_{i \in M(u)}^{} \sum_{l=1 }^{l=z}|C_i^{l, F}| + \sum_{i \in M(u)}^{} \sum_{l=1}^{l=z} \sum_{j=1}^{j=y_i^l} x_i^{l, P_j} \geq$ $demand(u)$, for each $u \in P.$
    \item $|N(v) \cap P| + x_i^{l, P_j} \geq |N[v] \setminus  P| -x_i^{l, P_j}-1$, for every $v \in C_i^{l, P_j}$.
    \item $x_i^{l, P_j} < l$, for each $i \in \{1,2,..., t\}$, $l \in \{1,2,..., z\}$ and $j \in \{1,2,..., y_i^l\}$.
    \end{itemize}
    }%
}\vspace{3mm} \\
In our ILP formulation, we have at most $\sum_{i = 1}^{t}\sum_{l = 1}^{z} y_i^l$ variables, where $z$ is the size of the largest clique outside the twin cover and $y_i^l \leq l-1$. The values of all the variables and the objective function are bounded by $n$. The constraints can be represented using $\mathcal{O}(\sum_{i = 1}^{t}\sum_{l = 1}^{z} y_i^l \cdot |T| \cdot log n)$ bits. With the help of Theorem 4, we will be able to solve the problem with the guess $(P, C_i^{l, F}$ and $C_i^{l, P})$ in FPT time. There are $2^{|T|}$ candidates for $P$ and there are $\sum_{i = 1}^{t}\sum_{l = 1}^{z} y_i^l \cdot \mathcal{O}(n)$ candidates for $(C_i^{l, F}$ and $C_i^{l, P})$. To obtain $S_C$, we solve $2^{|T|} \cdot \sum_{i = 1}^{t}\sum_{l = 1}^{z} y_i^l \cdot \mathcal{O}(n)$ ILP formulas, where each formula can be computed in $\mathcal{O}^*(f(|TC|, z))$ time. This concludes the proof of Theorem 7. \vspace{3mm} \\
\noindent \textbf{Theorem 8.} \textit{The} D\scriptsize{EFENSIVE ALLIANCE }\normalsize \textit{problem is fixed-parameter tractable parameterized by twin cover}. \vspace{3mm} \\
\textbf{Lemma 12.} If there exists a clique $C^* \in C_i$ of size at least $|N(C_i) \cap T|$ then there exist a defensive alliance $S \subseteq C^*$. \vspace{3mm} \\
\textit{Proof.} Consider a clique set $C_i$ outside the twin cover. Let $C^*$ be the clique from $C_i$ of size at least $|N(C_i) \cap T|$. Let $v$ be the vertex from $C^*$ that is also a part of $S$. The closed neighbourhood of $v$ is $|C^*|+|N(C_i) \cap T|$. In order to protect $v$, we need to push at least $\frac{|C^*|+|N(C_i) \cap T|}{2}$ neighbours of $v$ to $S$. This can be done in multiple ways, but we obtain a defensive alliance with smaller cardinality by picking all the $\frac{|C^*|+|N(C_i) \cap T|}{2}$ vertices from $C^*$ itself. This concludes that, if there is any clique $C^*$ of size at least $|N(C_i) \cap T|$ has a vertex in $S$ then there exists a defensive alliance, $S \subseteq C^*$. \qed \vspace{2mm}
We consider two cases: (1) There exists a vertex from a clique $C^* \in C_i$ of size at least $|N(C_i) \cap T|$ that is part of $S$. (2) No vertex from clique $C^* \in C_i$ of size at least $|N(C_i) \cap T|$ is a part of $S$. \vspace{2mm} \\
\indent For case 1, we solve $2^{|TC|}$ subproblems corresponding to each clique set. We find the optimal solution in each subproblem by considering a vertex from the shortest clique of size at least $|N(C_i) \cap S|$ to be in $S$. From Lemma 12, each subproblem can be solved in linear time in $n$. The minimum value among all the $2^{|TC|}$ subproblems will be the optimal solution for this case.\vspace{2mm} \\
\indent For case 2, we remove all the cliques from $G$ that cannot be a part of the solution and obtain a new graph $G'$. For each vertex $u \in P$, the value of $demand(u)$ will remain the same as calculated in $G$. The size of the largest clique outside the twin cover in $G'$ is at most $|TC|$. With the help of ILP formulation given in Theorem 7 and a bound on the largest clique outside the twin cover $z \leq |TC|$, the instance can be solved in FPT time. \vspace{2mm} \\
\indent The optimal solution to the problem is the minimum value obtained between the two cases. This concludes the proof of Theorem 8.
\section{Conclusions and Open Problems}
In this work, we have proved that the D\scriptsize{EFENSIVE ALLIANCE }\normalsize problem is \textit{polynomial-time solvable} on graphs with maximum degree at most 5 and NP-Complete on graphs with maximum degree 6. The byproduct of our result is that the problem is para-NP-hard parameterized by the maximum degree of the input graph. Therefore, one could also work on larger structural parameters than the maximum degree, such as the bandwidth and the maximum leaf number. A study of the offensive and powerful alliance problems on bounded degree graphs can also be considered. \vspace{3mm} \\
\indent We have also proved that the problem is fixed-parameter tractable parameterized by twin cover and distance to clique. The problem remains unsolved for the parameters modular width, and distance to cluster which is also an interesting direction to pursue.  It is interesting to study the parameterized complexity of the offensive and powerful alliances for the parameter twin cover and distance to cluster.

\section*{Acknowledgements}
The authors thank the anonymous reviewers for their valuable comments and suggestions.

%
%
%
\bibliographystyle{splncs04}
\bibliography{reference}

\end{document}